\documentclass[journal]{IEEEtran}
\usepackage{amsmath,amssymb,amsfonts,mathrsfs,mathtools,bm, bbm, dsfont,mathrsfs,amsthm,blkarray}
\usepackage[dvipdfm]{graphicx}
\usepackage[dvipsnames]{xcolor}

\usepackage{epsfig,epsf,psfrag,latexsym}

\usepackage{booktabs}
\usepackage{multirow,multicol}

\usepackage{mathtools,amssymb,lipsum}

\usepackage{cuted}
\usepackage{enumitem}

\usepackage[breaklinks,colorlinks, linkcolor=MidnightBlue, anchorcolor=MidnightBlue, citecolor=MidnightBlue, urlcolor=MidnightBlue]{hyperref}
\usepackage[hyphenbreaks]{breakurl}
\usepackage{cite}
\usepackage{enumitem}

 \def\old#1{}    % Please don't remove this... This command includes the text to be deleted.

\def\nn{\nonumber}
\def\beq{\begin{equation}}
\def\eeq{\end{equation}}
\def\bea{\begin{eqnarray}}
\def\eea{\end{eqnarray}}
\def\ba{\begin{array}}
\def\ea{\end{array}}

\def\bitem{\begin{itemize}}
\def\eitem{\end{itemize}}
\def\ben{\begin{enumerate}}
\def\een{\end{enumerate}}

\def\ie{{\it i.e.,\ \/}}

\def\alphabf{\hbox{\boldmath$\alpha$\unboldmath}}
\def\betabf{\hbox{\boldmath$\beta$\unboldmath}}

\def\etabf{\hbox{\boldmath$\eta$\unboldmath}}

\def\lambdabf{\hbox{\boldmath$\lambda$\unboldmath}}
\def\mubf{\hbox{\boldmath$\mu$\unboldmath}}

\def\omegabf{\hbox{\boldmath$\omega$\unboldmath}}

\def\Lambdabf{\mbox{$ \bf \Lambda $}}

\def\fbf{{\bm f}}

\def\pbf{{\bm p}}

\def\rbf{{\bm r}}

\def\ubf{{\bm u}}
\def\vbf{{\bm v}}

\def\xbf{{\bm x}}
\def\ybf{{\bm y}}

\def\rbf{{\bm r}}
\def\xbf{{\bm x}}
\def\ybf{{\bm y}}
\def\Abf{{\bm A}}

\def\Cbf{{\bm C}}
\def\Dbf{{\bm D}}

\def\Lbf{{\bm L}}

\def\Pbf{{\bm P}}

\def\Sbf{{\bm S}}

\def\Ec{{\cal E}}

\def\Gc{{\cal G}}

\def\Nc{{\cal N}}

\newcommand{\beqa}{\begin{eqnarray}}
\newcommand{\eeqa}{\end{eqnarray}}
\newcommand{\beqan}{\begin{eqnarray*}}
\newcommand{\eeqan}{\end{eqnarray*}}

\newcommand{\diag}{\mathop{\mathrm{diag}}}

\newcommand\T{{\mathpalette\raiseT\intercal}}
\newcommand\raiseT[2]{\raisebox{0.25ex}{$#1#2$}
}

\newcommand{\Rset}{\mathbb{R}}

\newcommand{\Gcal}{{\cal G}}

\newcommand{\Pcal}{{\cal P}}

\newcommand{\Wcal}{{\cal W}}

\renewcommand{\v}[1]{{\bm{#1}}}

\newcommand{\ol}[1]{\ensuremath{\overline{{#1}}}}
\newcommand{\ul}[1]{\ensuremath{\underline{{#1}}}}

\newcounter{l1}
\newcounter{l2}
\newcounter{l3}
\newcommand{\bdotlist}{\begin{list}{$\bullet$}{}}
\newcommand{\bboxlist}{\begin{list}{$\Box$}{}}
\newcommand{\bbboxlist}{\begin{list}{\raisebox{.005in}{{\tiny
$\blacksquare$ \ \ }}}{}}
\newcommand{\bdashlist}{\begin{list}{$-$}{} }
\newcommand{\blist}{\begin{list}{}{} }
\newcommand{\barablist}{\begin{list}{\arabic{l1}}{\usecounter{l1}}}
\newcommand{\balphlist}{\begin{list}{(\alph{l2})}{\usecounter{l2}}}
\newcommand{\bAlphlist}{\begin{list}{\Alph{l2}.}{\usecounter{l2}}}
\newcommand{\bdiamlist}{\begin{list}{$\diamond$}{}}
\newcommand{\bromalist}{\begin{list}{(\roman{l3})}{\usecounter{l3}}}

\newtheorem{lemma}{Lemma}
\newtheorem{proposition}{Proposition}

\newtheorem{definition}{Definition}

\title{Wholesale Market Participation of DERAs: \\ DSO-DERA-ISO Coordination}
\author{Cong Chen%~\IEEEmembership{Student Member,~IEEE,}
\quad
Subhonmesh Bose
\quad 
Timothy D. Mount
\quad Lang~Tong%~\IEEEmembership{Fellow,~IEEE}
\thanks{\scriptsize Part of the work was accepted by the 2023 IEEE Power \& Energy Society General Meeting  (PESGM) \cite{ChenBoseTong22Access}.}
\thanks{\scriptsize
Cong Chen and Lang Tong (\{cc2662, lt35\}@cornell.edu) are with the School of Electrical and Computer Engineering, Cornell University, Ithaca NY, USA. Subhonmesh Bose (boses@illinois.edu) is with the Department of Electrical and Computer Engineering at the University of Illinois Urbana-Champaign (UIUC), Urbana IL, USA.  Timothy Douglas Mount (tdm2@cornell.edu) is with the Dyson School of Applied Economics and Management, Cornell University, USA.}
\thanks{\scriptsize This work was supported in part by the National Science Foundation under Award 2218110 and 2038775 and the Power Systems and Engineering Research Center (PSERC) Research Project M-46.}
}

\begin{document}
\maketitle

\begin{abstract}
Distributed energy resource aggregators (DERAs) must share the distribution network together with the distribution utility in order to participate in the wholesale electricity markets that are operated by independent system operators (ISOs). We propose a forward auction that a distribution system operator (DSO) can utilize to allocate distribution network access limits to DERAs. As long as the DERAs operate within their acquired limits, these limits define operating envelopes that guarantee distribution network security, thus defining a mechanism that requires no real-time intervention from the DSOs for DERAs to participate in the wholesale markets. Our auctions take the form of robust and risk-sensitive markets with bids/offers from DERAs and utility's operational costs. Properties of the proposed auction, e.g., resulting surpluses of DSO and the DERAs, and the auction prices, along with empirical performance studies, are presented.

\end{abstract}


\vspace{-0.2in}
\section{Introduction}\label{sec:Intro}
The landmark ruling of the Federal Energy Regulatory Commission (FERC) Order 2222 in \cite{FERC20} aims to remove barriers to the direct participation of distributed energy resource aggregators (DERAs) in the wholesale market operated by independent system operators (ISOs) (or regional transmission operators). Since distributed energy resources (DERs) originate in  a distribution network, aggregated DERs must pass through the distribution grid managed by a distribution system operator (DSO) that can be the distribution utility or an independent entity. A coordination mechanism among the DSO, ISO, and DERAs is necessary to ensure system reliability and open access to all DERAs. FERC Order 2222 recognizes the significance of DSO-DERA-ISO coordination while leaving the specifics of the coordination design to the regulators, market operators, and stakeholders. 

DSO-DERA-ISO coordination poses significant theoretical and practical challenges. Net power injections from DERAs will likely depend on wholesale market conditions such as wholesale locational marginal prices (LMPs), real-time regulation service needs, and available behind-the-meter DERs in the distribution system. Notwithstanding these uncertainties, the DSO must ensure the reliable operation of the distribution grid, both in delivering services to all customers and allowing DERAs to offer services to the wholesale market. Moreover, any coordination mechanism must provide open and nondiscriminatory access to multiple competing DERAs operating over the same distribution network. 
%Equal access in this context means that the mechanism must not discriminate among the DERAs in how they participate or are compensated.

%DSO-DERA coordination poses significant theoretical, practical, and economic challenges. Power injections and withdrawals from DERs will likely depend on the wholesale market condition (such as locational marginal prices and real-time needs for regulation services) and the available resources (e.g.,  behind-the-meter renewables). Yet, with DER and wholesale price uncertainties, the DSO must ensure the safe operation of the distribution grid to reliably deliver power to all customers of the distribution utility and the DERAs. Any coordination mechanism that the DSO adopts must provide open and equitable access to multiple competing DERAs operating over the same distribution network. By equitable access in this context, we mean that the  mechanism cannot discriminate among the DERAs in how they participate or are compensated. 

DSO-DERA-ISO coordination has been actively debated since the release of FERC Order 2222. In \cite{Renjit:22EPRIReport}, coordination models have been classified into four categories, ranging from the least to the most DSO involvement. Type I models assume no DSO control (see, e.g.,  \cite{Alshehri&etal:20TPS,Gao&Alshehri&Birge:22}), because installed DER capacities are deemed to lie within the network's hosting capacity limits. One approach is to impose strict net injection limits \cite{AusNet17DER, SwitchDin23OE}  on individual prosumers so that the system's reliability is ensured as long as the limits are respected. In Type II models, e.g., \cite{NazirAlmassalkhi22TPS, LiuOchoaElta21PEM, LiuOchoaElta21TSG}, the DSO strives to prevent constraint violations, considering the randomness of power injections from DERs. Type III models involve coordination among DERAs, DSO, and ISO, where DERAs can provide distribution grid services in addition to delivering wholesale market products (see, e.g.,\cite{Papalex21DLMP, OpusOne21DOE}). Type IV models require DERA aggregation through the DSO, with the DSO performing all reliability functions and participating in the wholesale market on behalf of the DERAs as in \cite{GirigoudarRoald22T64, chen2019aggregate, MousaviWu22TPS, KonstantinosEta20TSG}. %\tcr{shall we explain in detail that \cite{GirigoudarRoald22T64, chen2019aggregate} is computing available flexibility for DSO-ISO coordination, and \cite{MousaviWu22TPS, Papalex21DLMP} is using DLMP. }

\begin{figure}[htbp]
    \centering
    \includegraphics[width=0.33\textwidth]{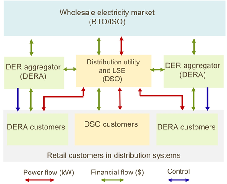}
    \caption{Power flow, financial flow, and control interactions in the  DSO-DERA-ISO coordination model.}
    \label{fig:DERAmodel}
\end{figure}
\vspace{-0.15in}

In this paper, we develop a Type II DSO-DERA-ISO coordination mechanism aimed at achieving efficient and reliable multi-DERA aggregations without significant deviations from existing DSO and ISO/RTO interaction models. Fig.~\ref{fig:DERAmodel} illustrates the power, financial, and control interactions among DSO, DERA, and ISO. Our coordination approach decouples the complex DSO-DERA-ISO interactions into nearly {independent, pairwise} interactions. In particular, we propose a  \emph{forward auction} run by the DSO that allows DERAs to acquire \emph{network access limits}--the right to inject or withdraw any amount of power within those limits over which the auction outcomes stand. These limits are auctioned off in a way that \emph{all} power transactions from DERAs within these limits will satisfy distribution system operation constraints. Thus, DERA's wholesale market interaction with the ISO in real-time can be agnostic to these operational constraints, and ISOs can dispatch DERs without direct visibility into the distribution network. The DERA-ISO interaction can include models such as demand response, virtual storage, etc., and \emph{any} way they choose to operate the DERs will not violate distribution network security as long as they remain within the network access limits they acquire from the DSO-operated auction. Thus, our design removes the need for DSO's interventions in ISO's real-time dispatch and DERAs' aggregation actions under normal system operating conditions.
% \tcb{Critical to realizing the real-time DSO-DERA-ISO decoupling is the proposed DSO-DERA interaction in a forward auction that allocates network access limits to DERAs to satisfy distribution system operation constraints.} %\tcb{In our continuing research \cite{ChenAlahmedMountTong23DERA}, we consider the aggregation strategies of a profit-seeking DERA that guarantees its customers no less surplus than they would have received from the regulated utility.}%  The impact of DERA aggregation strategies on wholesale market efficiency is also considered. 

We consider two network access allocation mechanisms. Sec.~\ref{sec:AccessRight} presents a robust optimization-based market clearing formulation for network access,  which guarantees satisfaction of network constraints when DERAs aggregate within their acquired access limits, thus removing the need for DSO to participate in real-time DERA-ISO interactions. DERAs do not even need knowledge of the underlying physical network when its aggregation strategy respects said limits.

The robust access allocation can be conservative in that it assumes the worst-case aggregation and network operating conditions.  We present a risk-based stochastic allocation mechanism in Sec.~\ref{sec:SOAccessRight} that allows DSO to share the common network resources subject to an acceptable risk constraint on network security violations. For both the robust and stochastic access limit auctions, we provide theoretical analyses and empirical studies to characterize and quantify access allocation properties, including the nonnegative surpluses for DSO, the benefits of participating in the access limit auction for DERAs, the  behavior of access allocation prices, and a comparison of the two formulations.

\vspace{-0.1in}
\begin{figure}[htbp]
    \centering
    \includegraphics[width=0.32\textwidth]{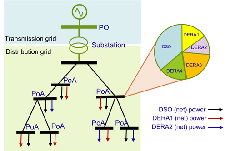}
    \caption{The distribution system and DER resources.}
    \label{fig:Distribution-DER}
\end{figure}
\vspace{-0.1in}
Our DSO-DERA coordination mechanism via access limits  shares both parallels and important differences with two distant relatives. One is the transmission right allocation problem for participants with bilateral contracts considered in the early years of wholesale market deregulation. Allocating physical transmission rights was deemed impractical and unnecessary (see \cite{Hogan92JRE_FTR}) with the ``loop-flow'' problem in meshed transmission networks. Ultimately, wholesale markets evolved to adopt a centrally coordinated economic dispatch run by the ISO, where bilateral transactions are protected from the risks of LMP fluctuations using derivatives such as financial transmission rights. In our proposed coordination mechanisms, no central authority coordinates DERAs in real-time. Making access limit allocation \emph{physical} allows DERAs to inject or withdraw any amount of power within their purchased access limits. The loop-flow problem does not affect our design, thanks to the radial nature of distribution networks. Also, we deliberately separate access limit allocation from real-time decisions; we envision the forward auction for access limits to run once a day or a week. We posit that tight coordination of dispatch decisions via a centralized market mechanism, matched to that operated by the ISO, may be impractical in the near-term and possibly unnecessary, owing to smaller trade volumes and less stringent system constraints in the distribution grid.

Second, our mechanism is reminiscent of the notion of operating envelopes defined by DSO-imposed net nodal DER injection and withdrawal limits. For instance, an Australian market imposes 5kW net-injection limits on residential customers with DERs \cite{AusNet17DER}. Instead of using pre-determining dynamic operating envelopes as in \cite{SwitchDin23OE, LiuOchoaElta21PEM, LiuOchoaElta21TSG}, we auction off these envelopes among competing DERAs based on each DERA's aggregation needs and aggregation strategy.

%Moreover, we propose an auction that allows open access to multiple DERAs to operate over a distribution network; this design is thematically different from time-varying injection and withdrawal ranges for individual DERs in \cite{LiuOchoaElta21PEM, LiuOchoaElta21TSG}.

% as they both construct the distribution network injection and withdrawal range, they  differ in allocation methods and imposed targets. The operating envelope can be static (eg. the 5kW injection operating envelope used in Australia \cite{AusNet17DER}) and dynamic (eg. the time-varying injection and withdrawal range in \cite{LiuOchoaElta21PEM, LiuOchoaElta21TSG}, which maximizes the network access limits while maintaining security constraints). Both types of operating envelopes are imposed on individual customers directly by the DSO, and it's unclear how DSO will interact with the aggregator about the operating envelopes. The robust access limit proposed in this paper is imposed on DERA by DSO, and we maximize the social welfare in the allocation model.

The paper is organized as follows. We begin with the preliminaries of the network and the DSO-DERA-ISO coordination models in Sec.~\ref{sec:A_Access}. Then, we present the robust and the stochastic network access allocation problems and their properties in Sec.~\ref{sec:AccessRight} and  Sec.~\ref{sec:SOAccessRight}, respectively. A simple yet illustrative example is discussed in Sec.~\ref{sec:example.4}. Then, we provide a numerical case study on a 141-bus distribution network in Sec.~\ref{sec:CaseStudies}. 
Theoretical claims are proven in the appendix.

\vspace{-0.15in}
{\small
\begin{table}[htbp]
\caption{Major symbols}\label{tab:symbols}
\begin{center}
\vspace{-1em}
\begin{tabular}{ll}
\hline
$\v{p}_0,\ol{\v{p}}_0, \ul{\v{p}}_0$& power injections and network accesses of DSO\\
$\v{p}_k,\ol{\v{C}}_k, \ul{\v{C}}_k$&  power injections and network accesses of DERA $k$\\
$\v{\ol{C}}_k^{\mbox{\tiny min}}, \v{\ul{C}}_k^{\mbox{\tiny min}}$& minimum injection/withdrawal access of DERA $k$\\
$\v{\ol{P}}^{\mbox{\tiny max}}, \v{\ul{P}}^{\mbox{\tiny max}}$&  injection/withdrawal access limits of the network\\
$\varphi_k$&  utility of DERA $k$, induced by its bid\\
$J$:&  operational cost of DSO\\
$\Abf$:& network parameters for  linearized power flow\\
% $\fbf$:&  power flow over all branches.\\
% $\vbf$:& voltage over all buses.\\
$\v{\ol{b}}, \v{\ul{b}}$ & limits on network voltage/power flows\\
$K, S$& total number of DERAs and scenarios \\
% $S$& total number of scenarios\\
\hline
\end{tabular}
\end{center}
\end{table}
}

% \section{DERA-DSO coordination}\label{sec:DERA_DSO}
% \input{DERA_DSO_v0}
\vspace{-0.2in}
\section{Network and Coordination Models}\label{sec:A_Access}
%The linearized security constraints for voltage and line flow limits of the distribution network are adopted in the network access allocation model for  the DSO-DERA coordination. 

%\subsection{Linearizede power flow model}\label{sec.Linflow}
Consider $K$ DERAs operating over a radial distribution network across $N$ buses shown in Fig.~\ref{fig:Distribution-DER}, where bus 1 is the reference bus. Let the power injection profile from DERA $k$ across the network be given by $\v{p}_k \in \Rset^{N}$, and the total DERA net injection to the system is given by $\sum_{k=1}^K \v{p}_k$. Let the power injection profile from the DSO's customers be given by $\v{p}_0 \in \Rset^N$. Assuming a uniform power factor for all power injections, the reactive power injection profile is given by $\alpha \left(\sum_{k=1}^K \v{p}_k+\v{p}_0\right)$. This assumption simplifies our presentation but is not crucial to the design of our auctions. These real and reactive power injections then induce power flows and voltage magnitudes over the distribution network that are related via the power flow equations. In this work, we adopt a linear distribution power flow (LinDistFlow) model  \cite{baranWu89TPDdistFlow} adopted from \cite{Low19PowerSystemAnalysis}. Voltage and line capacity/thermal limits define security constraints. As explained in Appendix \ref{sec:MLPF}, these constraints take the form,
% The network constraint is explained by the vector of power flows over the distribution lines and the squared voltage magnitudes over the buses, excluding the reference bus via a linear relationship, 
% $\v{A} \left(\sum_{k=1}^K \v{p}_k+\v{p}_0\right)$, that is then constrained as }
% \begin{align}
\beq
    \ul{\v{b}} \leq \v{A} \left(\sum_{k=1}^K \v{p}_k+\v{p}_0\right) \leq \ol{\v{b}}.
    \label{eq:volLBUB}
\eeq
% \end{align}
Inequalities are interpreted element-wise.
% \footnote{Note that other linearized power flow models can also be applied here.}
% \footnote{Ignoring the reference bus, we have $\v{v} \in \Rset^{N-1}, \ol{\v{v}} \in \Rset^{N-1},  \ul{\v{v}} \in \Rset^{N-1}$.
% \begin{align}
% \begin{bmatrix}
% \v{f} \\ \v{v} \end{bmatrix}
%  = \begin{bmatrix} 
%  \v{A}_{fp}  & \v{A}_{fq}
%  \\
%  \v{A}_{vp} & \v{A}_{vq}
%  \end{bmatrix} 
%  \begin{bmatrix} \v{p} + \v{p}_0 \\ \alpha (\v{p} + \v{p}_0)\end{bmatrix} 
%  = \v{A}(\v{p} + \v{p}_0). \label{eq:lineA}
% \end{align}

% The detailed derivation of $\v{A}, \ol{\bbf}, \ul{\bbf}$ is relegated to Appendix \ref{sec:MLPF} to maintain the continuity of exposition.
% for the LinDistFlow model is included in the appendix together with a 5-bus example. The power flows and the voltage magnitudes must remain within the engineering limits of the network, \ie
% \begin{align}\label{eq:volLBUB}
   % \v{\ul{b}}  \leq (\v{f}, \v{v}) \leq   \v{\ol{b}}.
% \end{align}

With the linearized  network model, we propose forward auctions for DSO-DERA coordination. In these auction models, the DERAs bid for injection/withdrawal \emph{access} at various buses of the distribution network, where the DERA commands DERs from its customers; see Fig.~\ref{fig:Distribution-DER}. In Sec.~\ref{sec:AccessRight} and  Sec.~\ref{sec:SOAccessRight}, we describe, respectively the robust and the stochastic, optimization problems to clear the forward auctions and the settlements for the DERAs. 
% These auctions utilize utility functions of DERAs that are inferred from bids/offers and the operational cost of the DSO.
The auctions determine the \emph{range} of injection and withdrawal access for each DERA at each bus of the network, and the DERAs' payments to acquire those access limits. We thus design the auction of an operating envelope, where the DERA can inject/withdraw \emph{any} amount of power from the DERs they command within this envelope that they purchase from the DSO. Thus, the real-time power of DERA stays within the limits/envelopes they purchase at the points of aggregation (PoAs) from the DSO through the proposed forward auctions. Our network model contains buses at the PoAs, but we do not model individual customers/DERs downstream from the PoAs.  Transactions between a DERA and the ISO occur at the  point of interconnection (PoI).

% is designed to auction off the network capacities at all buses in the distribution grid. That way, DERAs can acquire the network accesses from DSO, and pay DSO for operation services. And all DERAs' bids into the wholesale market will be revealed at the  point of interconnection (POI) to provide services for the transmission grid. To establish a proper granularity of the distribution network, we introduce the point of aggregation (PoA). PoA corresponds to the main buses in the distribution network, which  have higher voltage levels than the average. In part II paper, we establish the DER aggregation model under PoA.

Our auctions for network access are hosted ahead of real-time operations--it can be a few hours to a week in advance of power delivery. To participate, all DERAs must submit their bids/offers to this forward auction and they are cleared simultaneously. Through this auction, DERAs get access to limits/capacities that they must obey during real-time operations. Since the auction outcome binds over multiple real-time interactions, we account for the range of possible operating conditions through robust and stochastic programming based auction-clearing formulations.

% Note that DERA can only update its bid into the real-time wholesale market about one hour ahead \cite{Schiro19}. Therefore DERA needs to acquire the distribution network access at least about one hour ahead to construct its bid in the wholesale market operated by ISO.} %The time granularity of how frequent this forward network access auction host is at most hourly-based. } 

\vspace{-0.05in}
\section{The Robust Auction Model}\label{sec:AccessRight}

% We consider a robust resource allocation mechanism where the DSO auctions off injection and withdrawal limits across the distribution network to the DERAs. Specifically, 
We now design an auction for network access that accounts for a variety of real-time operating conditions through a robust optimization formulation. Specifically, let $\ul{\v{C}}_k \in \Rset^{N}$ and $\ol{\v{C}}_k \in \Rset^{N}$ denote the vectors of (real) power withdrawal and injection capacity limits acquired by DERA $k$. Then, all power injections from assets controlled by DERA $k$ must respect
$\v{p}_k \in [-\v{\ul{C}}_k, \v{\ol{C}}_k]$. Let the DSO's own customers have net power injections $\v{p}_0$ that take values in the set $[-\v{\ul{p}}_0, \v{\ol{p}}_0]$. Given these ranges of the various power injections, the DSO solves, 
\begin{subequations}
    \begin{align}
    & \underset{\v{\ul{C}}, \v{\ol{C}}, \v{\ol{P}}, \v{\ul{P}}}{\text{maximize}} && \sum_{k=1}^K \varphi_k(\v{\ul{C}}_k,\v{\ol{C}}_k)-J(\ol{\Pbf},\ul{\Pbf}), \label{eq:objSS}
    \\
    & \text{subject to} 
    &&\v{\ol{C}}_k^{\mbox{\tiny min}} \le \v{\ol{C}}_k , ~~\v{\ul{C}}_k^{\mbox{\tiny min}} \le \v{\ul{C}}_k, 
    \label{eq:ROC}
    \\
    &&&  \v{\ol{P}} \le \v{\ol{P}}^{\mbox{\tiny max}}, ~~\v{\ul{P}} \le \v{\ul{P}}^{\mbox{\tiny max}}, \label{eq:ROP}
    \\
    &&& \v{\ol{P}} = \sum_{k=1}^K \v{\ol{C}}_k + \v{\ol{p}}_0,  \label{eq:Rol}
    \\
    &&&
    \v{\ul{P}} = \sum_{k=1}^K \v{\ul{C}}_k + \v{\ul{p}}_0, \label{eq:Rul}
    \\ 
%    &&& 
%    \v{p} =\sum_{k=1}^K \v{p}_k,
%    \\
    &&&
    \v{\ul{b}}
    \leq \v{A}\left(\sum_{k=1}^K \v{p}_k+ \v{p}_0 \right) 
    \leq 
    \v{\ol{b}},\label{eq:ROline}
    \\
    &&& \text{for all } \v{p}_k \in [-\v{\ul{C}}_k, \v{\ol{C}}_k], \v{p}_0 \in [-\v{\ul{p}}_0, \v{\ol{p}}_0],\label{eq:RObd}
    \\
    &&& \text{for } k = 1,\ldots,K. \notag
    \end{align}
    \label{eq:auction}
\end{subequations}
Here, DERA $k$ provides the bid $\varphi_k:\Rset^{2N} \rightarrow \Rset$ to the DSO, where $\varphi_k(\underline{\Cbf}_k,\overline{\Cbf}_k)$ represents DERA $k$'s willingness to pay for power transactions. Bid/offer construction for $\varphi_k$ depends on the DERA's aggregation strategy; we refer to our work in \cite{ChenAlahmedMountTong23DERA}  for a candidate construction. Let $\ol{\Pbf}\in \Rset^{N}$ and $\ul{\Pbf}\in \Rset^{N}$ represent the vectors of the total injection and withdrawal capacities, respectively. Also, define $\v{\ol{C}}:=(\v{\ol{C}}_k),\v{\ul{C}}:=(\v{\ul{C}}_k)$ as the matrices that collect the access limits across the $K$ DERAs. The operational cost of the DSO is encoded in  $J:\Rset^{2N} \rightarrow \Rset$ that is assumed to be convex and non-decreasing. Being a regulated monopoly running the distribution grid, we anticipate that $J$ will include the cost of reactive power support, network maintenance, line losses, etc. required to maintain quality of service to existing retail customers. See \cite{Yeddanapudi08Maintainance}  for example cost component constructions.
% is under regulation in this  network access auction because DSO is a natural monopoly over the distribution network. The current operation cost of DSO can be recovered by DSO from the payment of its customers. This provides a benchmark of DSO to construct its bid-in cost and charge DERA for the network access.
The objective function then represents the induced social surplus from DERAs' bids and DSO's costs.
% \beq\label{eq:SSRO}
% {\Scal}(\v{\ul{C}}, \v{\ol{C}}, \v{\ol{P}}, \v{\ul{P}}):=\sum_{k=1}^K\varphi_k(\v{\ul{C}}_k,\v{\ol{C}}_k)-J(\ol{\Pbf},\ul{\Pbf}).
% \eeq
Let $\varphi_k$ be concave and non-decreasing for each $k$.
Additionally, $(\v{\ol{C}}_k^{\mbox{\tiny min}}, \v{\ul{C}}_k^{\mbox{\tiny min}})$ are the vectors of minimum injection and withdrawal capacities that DERA $k$ is willing to acquire across the network, and $(\v{\ol{P}}^{\mbox{\tiny max}}, \v{\ul{P}}^{\mbox{\tiny max}})$ are the vectors of  the maximum  injection and withdrawal capacities across the network. All capacity limits are assumed nonnegative.  Equations \eqref{eq:ROC} and \eqref{eq:ROP} encode the DERAs' minimum access requirements and the DSO's maximum access limits, respectively. Equations \eqref{eq:Rol} and \eqref{eq:Rul} define the total injection and withdrawal accesses in terms of DSO's access limits and those sold to individual DERAs. With the linearized network model in \eqref{eq:volLBUB}, the relations in \eqref{eq:ROline} and \eqref{eq:RObd} enforce the engineering constraints of the grid for \emph{every possible} power injection profile from the DSO's customers and those of all DERAs within their acquired capacities. 
% \tcb{Note that the bid-in cost $J$ of DSO is under regulation in this  network access auction because DSO is a natural monopoly over the distribution network. The current operation cost of DSO can be recovered by DSO from the payment of its customers. This provides a benchmark of DSO to construct its bid-in cost and charge DERA for the network access.}

%  The assumption below guarantees a convex optimization.%allocation.% mechanism. 
% \begin{assumption}\label{assump:CVX}
%     DSO cost function $J$ is differentiable and convex. Bid-in benefit function of DERAs, \ie $\varphi_k, \forall k$, are differentiable, concave and non-decreasing.% with $J({\bf 0},{\bf 0})=0$.  with $\varphi_k(\v{0},\v{0})\geq 0$
% \end{assumption}

%In addition to the advantage of the global optimal solution and less computation burden, the convex  allocation mechanism  also brings nice properties related to pricing and individual rationality, which are explained in Sec.~\ref{sec:LMAP-R}.

The formulation in \eqref{eq:auction} contains robust constraint enforcement in \eqref{eq:ROline}-\eqref{eq:RObd}, but it admits a reformulation as a classic convex program that we present next. The result requires additional notation. For any scalar $z$, define $z_+ := \max\{z,0\}$ and $z_- := z_+ - z$. Define the same for a matrix/vector, where the operations are carried element-wise.

\begin{lemma}\label{lemma:EqRO}
Problem \eqref{eq:auction} is equivalent to \begin{subequations}
    \begin{align}
    & \underset{\v{\ul{C}}, \v{\ol{C}}, \v{\ol{P}}, \v{\ul{P}}}{\text{\mbox{maximize}}} && \sum_{k=1}^K\varphi_k(\v{\ul{C}}_k,\v{\ol{C}}_k)-J(\ol{\Pbf},\ul{\Pbf}), \label{eq: ROOBJ}
    \\
    & \text{\mbox{subject to} }
    &&   \text{ for } k =1, \ldots, K, \nn\\
    &~~\ol{\etabf},\ul{\etabf}:&&\v{\ol{C}}_k^{\mbox{\tiny min}} \le \v{\ol{C}}_k , ~~\v{\ul{C}}_k^{\mbox{\tiny min}} \le \v{\ul{C}}_k , 
    \label{eq:auction.2.Cmin}
    \\
    &~~\ol{\omegabf},\ul{\omegabf}:&&  \v{\ol{P}} \le \v{\ol{P}}^{\mbox{\tiny max}}, ~~\v{\ul{P}} \le \v{\ul{P}}^{\mbox{\tiny max}}, \label{eq: ROPbd}
    \\
    &~~~~~~~~\ol{\lambdabf}:&& \v{\ol{P}} = \sum_{k=1}^K \v{\ol{C}}_k + \v{\ol{p}}_0,
    \\
    &~~~~~~~~\ul{\lambdabf}:&&
    \v{\ul{P}} = \sum_{k=1}^K \v{\ul{C}}_k + \v{\ul{p}}_0,
    \\
    &~~~~~~~~\ol{\mubf}:&&
    \v{A}_+ \v{\ol{P}} + \v{A}_-  \v{\ul{P}}
    \leq 
    \v{\ol{b}}, \label{eq:ROol}
    \\
    &~~~~~~~~\ul{\mubf}:&&
    \v{\ul{b}} \leq -\v{A}_+  \v{\ul{P}} - \v{A}_- \v{\ol{P}}.\label{eq:ROul}
    \end{align}
    \label{eq:auction.2}
\end{subequations}
\end{lemma}
% \vspace{-0.2in}
We remark that for the linear power flow model developed in Appendix \ref{sec:MLPF}, $\v{A} = \v{A}_+$ and $\v{A}_- = \mathbf{0}$. Our auction mechanism and its properties hold more generally for any linear approximation to power flow equations, and hence, we present our results with a more general $\v{A}$.
Associate Lagrange multipliers with the various constraints in \eqref{eq:auction.2} as shown. We now introduce the prices that will define the settlements with the DERAs using the \emph{optimal} Lagrange multipliers (indicated with $\star$) for \eqref{eq:auction.2}.
% Equations \eqref{eq:ROul} and \eqref{eq:ROol} correspond to the worst-case scenarios for the original constraints \eqref{eq:ROline} and \eqref{eq:RObd}.

 % \subsection{The Locational Marginal Access Price}\label{sec:LMAP-R}

\begin{definition}[LMAP-R]
For the robust access allocation problem \eqref{eq:auction.2}, the locational marginal access price for access limits to the distribution network is defined by the vector of optimal dual solutions $\ol{\lambdabf}^\star \in \Rset^N, \ul{\lambdabf}^\star \in \Rset^N$ of \eqref{eq:auction.2}. 
\end{definition}
Specifically, the injection and the withdrawal access price at bus $i$ are given by $\ol{\lambda}^{(i)}$ and $\ul{\lambda}^{(i)}$, respectively. With these prices, for obtaining injection and withdrawal access of $\ol{\v{C}}_k$ and $\ul{\v{C}}_k$, respectively, DERA $k$ pays
\begin{align}
    \Pcal_k(\v{\ol{C}}_k^{\star}, \v{\ul{C}}_k^{\star})=\ol{\lambdabf}^{\star\intercal} \v{\ol{C}}_k^{\star} + \ul{\lambdabf}^{\star\intercal} \v{\ul{C}}_k^{\star}
\end{align}
to the DSO.
%Optimal dual solutions  of (\ref{eq:auction.2}), respectively, define the vector of locational injection and withdrawal allocation prices for DERAs at different buses.
Having derived the prices from an auction that is reminiscent of the economic dispatch problem solved by RTO/ISO in wholesale electricity markets, LMAP-R shares strong parallels with locational marginal prices (LMPs). For example, LMAP-R is nodally uniform. That is, all DERAs pay the same injection and withdrawal access price at a specific distribution bus. These prices, however, may differ across locations in the distribution network. 

The second parallel between LMAP-R and LMP comes from the interpretations of these prices as sensitivities of the optimal objective function value of the market clearing problem to nodal parameters. For LMPs, the price of a bus equals the sensitivity of the optimal power procurement costs to nodal demands. For LMAP-R, it is the sensitivity of the induced social welfare $\Wcal$ to DSO's own access limits $\ol{\v{p}}_0$, $\ul{\v{p}}_0$. More precisely, define the optimal social welfare from the optimal value of \eqref{eq:auction.2} as $\Wcal^\star(\ol{\v{p}}_0, \ul{\v{p}}_0)$. Then, envelope theorem, per \cite[Chapter 7]{still18lectures}, states that when $\Wcal^\star$ is differentiable,
\begin{align}
    \ol{\lambdabf}^\star 
    =  -\nabla_{\ol{\v{p}}_0}\Wcal^\star(\ol{\v{p}}_0, \ul{\v{p}}_0),
    \;
    \ul{\lambdabf}^\star 
    = 
    -\nabla_{\ul{\v{p}}_0} \Wcal^\star(\ol{\v{p}}_0, \ul{\v{p}}_0),
\end{align}
which represents the marginal decrease in social welfare when supporting network access by DSO rather than selling access to DERAs. We further shed light on the contributions of network parameters to LMAP-Rs in our next result. 
\begin{proposition}\label{Prop:LAP}
LMAP-R satisfies
\begin{align}
\begin{aligned}
    \ol{\lambdabf}^\star = \nabla_{\ol{\Pbf}}J(\ol{\Pbf}^\star,\ul{\Pbf}^\star)+\Abf_+^{\T}\ol{\mubf}^\star + \Abf_-^{\T}\ul{\mubf}^\star+\ol{\omegabf}^\star,
    \\
    \ul{\lambdabf}^\star=\nabla_{\ul{\Pbf}}J(\ol{\Pbf}^\star,\ul{\Pbf}^\star)+\Abf_-^{\T}\ol{\mubf}^\star+\Abf_+^{\T}\ul{\mubf}^\star+\ul{\omegabf}^\star.
\end{aligned}
\label{eq:LMAPR.sensitivity}
\end{align}
\end{proposition}
% By envelope theorem and KKT conditions of \eqref{eq:auction}, t
The first term in \eqref{eq:LMAPR.sensitivity} equals the DSO's marginal cost for disseminating access limits at the optimum of \eqref{eq:auction.2}. Our result then characterizes the price markup in LMAP-R above said marginal cost. Specifically, if voltage and line constraints are non-binding, and the access allocations do not reach the injection limits, the remaining terms in $\ol{\lambdabf}^\star, \ul{\lambdabf}^\star$ vanish. In addition, if DSO's operational cost structure is additive and uniform across the buses of the distribution network, then LMAP-Rs become equal at all buses. The tighter the constraints on total access limits being auctioned (encoded in the entries of $\v{\ol{P}}^{\mbox{\tiny max}}$, $\v{\ul{P}}^{\mbox{\tiny max}}$) are, and the more stringent the network constraints  (represented in the entries of $\ol{\v{b}}$, $\ul{\v{b}}$) are, we expect LMAP-Rs to differ from DSO's marginal costs.

Next, we investigate DSO's and the DERAs' surplus at an optimal robust access allocation. Define DSO's surplus as
\begin{align}
\hspace{-0.13in}
% \begin{aligned}
        \Pi^{\mbox{\tiny DSO}}
        :=\sum_{k=1}^K\Pcal_k(\v{\ol{C}}_k^{\star}, \v{\ul{C}}_k^{\star}) - \left( J(\ol{\Pbf}^\star,\ul{\Pbf}^\star) -J(\v{\ol{p}}_0,\v{\ul{p}}_0)
    \right).
% \end{aligned}
    \label{eq:surplus.DSO}
\end{align}
The first term equals the total rent that the DSO collects from the DERAs. The second summand equals the cost that the DSO affords when allowing DERAs to operate over the distribution network. Specifically, $J(\ol{\Pbf}^\star,\ul{\Pbf}^\star)$ equals the cost of the net injection and withdrawal access $\ol{\Pbf}^\star,\ul{\Pbf}^\star$ when the DSO provides the network accesses for DERAs and itself, and $J(\v{\ol{p}}_0,\v{\ul{p}}_0)$ equals the operational cost for access required to support the DSO's customers alone. Thus, the last summand in \eqref{eq:surplus.DSO} measures the cost of the DSO when supporting the network accesses to the DERAs. Next, define DERA $k$'s surplus as
\begin{align}
    \Pi_k^{\mbox{\tiny DERA}}:= \varphi_k(\v{\ul{C}}_k,\v{\ol{C}}_k) - \Pcal_k(\v{\ol{C}}_k^{\star}, \v{\ul{C}}_k^{\star}),
    \label{eq:Pi.DERA.def}
\end{align}
which equals the induced utility (inferred from the bid) minus the payment to the DSO.

% Can the LMAP-based settlement cover DSO's operating cost and maintain a positive surplus for DERAs?  The answer is affirmative by Proposition~\ref{Prop:RA}. 

\begin{proposition}\label{Prop:RA} 
(i) $\Pi^{\mbox{\tiny DSO}} \geq 0$, (ii) $\Pi_k^{\mbox{\tiny DERA}} \geq 0$, if $\varphi_k(\v{0},\v{0})\geq 0$ and one of the following two conditions holds: (a) % \begin{itemize}
    % \item  
    $\v{\ol{C}}_k^{\mbox{\tiny min}}={\bf 0}, \v{\ul{C}}_k^{\mbox{\tiny min}}={\bf 0}$, or (b)
    % \item  
    the constraints in \eqref{eq:auction.2.Cmin} are non-binding at optimality, i.e.,  $\v{\ol{C}}_k^\star > \v{\ol{C}}_k^{\mbox{\tiny min}}$,$\v{\ul{C}}_k^\star >  \v{\ul{C}}_k^{\mbox{\tiny min}}$.
% \end{itemize}
\end{proposition}
The last result suggests that DSO always gains from running the auction in that its surplus is nonnegative. For DERAs, nonnegative surplus is assured under certain sufficient conditions. Among these conditions, $\varphi_k(\v{0},\v{0})\geq 0$ is natural as one expects the inferred utility of DERA to be nonnegative for any nonnegative value of the access limits. 

The condition $\ol{\Cbf}_k=\ul{\Cbf}_k=0$ indicates that DERA $k$ may obtain zero network access limits at some buses. Such a condition is violated when the DERA may require a minimum  demand to be met or net injection cannot be curtailed beyond a threshold. In such an event, the DERA may need side-payments to make the auction outcome favorable for its participation. The design and discussion of side-payments are relegated to a future effort. We remark that this phenomenon is reminiscent of the challenge in wholesale markets where minimum generation constraints can negate a generator's rationale for market participation. Additionally, we expect the lower bounds $\v{\ol{C}}_k^{\mbox{\tiny min}}, \v{\ul{C}}_k^{\mbox{\tiny min}}$ to be non-binding for DERAs that are \emph{marginal} to the auction, i.e., when their marginal implied costs define the prices. 
% are marginal market participants, i.e., \footnote{The marginal market participants take the same definition as marginal generators in the electricity market}.

% are positive, is analogous to LMP when generators have minimum generation limits.   In that case, the individual rationality of LMP is lost.  Addressing this issue may require the DERA to internalize the possible losses and structure its bids accordingly. Out-of-the-market action is another possible solution, which is outside the scope of this work. 

Our next result sheds light on how LMAP-Rs behave along the distribution feeder. Unlike our last two results, the next one specifically utilizes the power flow model presented in Appendix \ref{sec:MLPF}. To present the result, we need additional notation. We say bus $n$ is an \emph{ancestor} of bus $m$ in the distribution network if $n$ lies on the unique (undirected) path from the reference bus to bus $m$.

% It is intuitive that aggregation at the end of the feeder may be more costly as the voltage and power flow constraints may be more pronounced. 

%In Part II paper, we propose a method to compute the minimum profitable network access requirement of the marginal DERA participating in this access allocation. However, DERA still faces the risk of a deficit when the access allocation price is unexpectedly high. In practice, DERA can hedge the price volatility for the minimum access requirement. Thus we ignore this part of the negative surplus in our analysis.

%Moreover, we observe that, along one distribution line from the root node to the leave node, the access allocation price is non-decreasing if providing uniform DSO marginal cost along the line. And the following proposition explains this phenomenon in detail.
\begin{proposition}\label{Prop:PM2}
When $J$ is linear and homogeneous across buses ($J(\ol{\Pbf},\ul{\Pbf}) = \sum_{i=1}^N  \ol{J} \cdot \ol{P}^{(i)} + \sum_{i=1}^N \ul{J} \cdot \ul{P}^{(i)} $) and the constraints in \eqref{eq: ROPbd} are non-binding at all buses, then $\ol{\lambda}^{(n)\star} \geq \ol{\lambda}^{(m)\star}$ and $\ul{\lambda}^{(n)\star}  \geq \ul{\lambda}^{(m)\star}$, if bus $n$ is an ancestor of bus $m$. 
\end{proposition}
Said differently, LMAP-Rs do not decrease along the network away from the substation under certain conditions. Our numerical experiments reveal that these conditions are only sufficient for price monotonicity; the monotonic trend holds even when these conditions are not met.  Such a price monotonicity reveals that it is costlier to guarantee the voltage limits and line capacity limits for the leaf buses, compared to those closer to the substation. The non-binding nature of the constraints is only \emph{sufficient} for the conclusion to hold; our simulations will show that they are \emph{not} necessary.
% The assumption that  constraints in \eqref{eq: ROPbd} are non-binding indicates that the distribution network mainly has binding constraints from voltage and line capacity limits.

% The above proposition indicates that the radial nature of the distribution grid often leads to increasing LMAPs from the PCC to leaf buses. (See numerical results). The homogeneous DSO operating cost can be relaxed to  $\frac{\partial J^\star(\cdot)}{\partial \ol{P}_n} \geq \frac{\partial J^\star(\cdot)}{\partial \ol{P}_m}$ and $\frac{\partial J^\star(\cdot)}{\partial \ul{P}_n} \geq \frac{\partial J^\star(\cdot)}{\partial \ul{P}_m}$, and the price monotonicity still holds.  

%Proof of the above proposition is shown in Sec.~\ref{sec:ProfProp4}, which follows the property of the LinDistFlow model parameterized by matrix $\Abf$. And it also holds for other linearized power flow models like logarithmic transform voltage magnitude (LTVM) linearization \cite{Li17TPSLTVM, molzahnHiskens19surveyRelaxPowerflow} and the state-independent linear power flow \cite{YangKangTPS16linearPF}. %Such a proposition indicates that the locational allocation prices is monotonically nondecreasing over a distribution feeder from the root to the leave, when the DSO operation cost is the same over the distribution network.

% marginal DSO cost function at bus $n$ is no less than that at bus $m$, locational allocation price at bus $n$ is no less than that at bus $m$

% \section{Bids and Offers from Profit-seeking DERAs }\label{sec:InjDERAopt}
% \input{ICaggregation_v5}

\section{The Stochastic Auction Design}\label{sec:SOAccessRight}
The robust access allocation requires the network constraints to be satisfied for \emph{all possible} injections from DERAs' and the DSO's customers. Naturally, the resulting allocations are dictated by the worst-case power injection/withdrawal combinations during the forward auction and fully ignore the statistics of network usage. Such an approach inherently limits the DERAs' collective network access. 
We present in this section a risk-based stochastic allocation mechanism that allows \emph{controlled violation} of the network constraints in the access allocation auction in a way that accounts for the statistics of power transactions by DSO's customers. Stochastic resource allocation with controlled violation is common in many areas, especially in computer systems and communication networks, where the resources are constrained and usage patterns are random, especially over time.
% \footnote{An example is the parking permit allocation when more permits are issued than the number of available spots because not all spots are used all the time.} 
Such an allocation builds on the premise that random usage patterns often do not overlap, and this time-multiplexing allows higher access limits to be accommodated than the roubust counterpart.
% The idea is to multiplex stochastic access (random access) while meeting a significantly higher demand than the strict partition of network capacity can accommodate.
Notice that we allow for possible controlled violations only in the forward auction. Said violations in the auction do not amount to actual violations in real time. In practice, DSO can implement mechanisms to curtail access to enforce reliability constraints in real-time; we side-step such considerations and  purely focus on the merits of a stochastic forward auction.

% Except for the optimization involved in access allocation, the definition of the stochastic locational marginal access price, herein referred to as LMAP-S, and its related properties are parallel to the robust allocation schemes.

%To avoid over conservativeness of the robust allocation in the last section, we provide a stochastic access allocation mechanism with conditional value at risk (CVaR) measuring the risk for the violation of the network operation constraints. In parallel, we derive properties like the revenue adequacy, incentive compatibility, and price monotonicity on expectation. 

% \subsection{Risk-Based  Stochastic  Access Allocation}

\begin{figure}[htbp]
    \centering
    \includegraphics[width=0.4\textwidth]{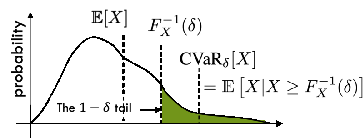}
    \caption{CVaR of a random variable $X$. }
    \label{fig:CVaRDef}
\end{figure}
\vspace{-0.15in}
When the power transactions by DSO's customers are considered random, the power flows and the bus voltages become random quantities. In the forward auction, we limit the \emph{risk} of constraint violations, where we measure this risk via the conditional value at risk (CVaR) measure. CVaR, analyzed and popularized by \cite{Rockafellar00CVaR,rockafellar2002conditional}, has recently been advocated in power system planning, e.g., in \cite{madavan22CVaR}. For a random variable (think loss) $X$ with smooth cumulative distribution function $F_X$, CVaR at level $\delta \in [0,1)$ equals the conditional mean of $X$ over the $(1-\delta)$-tail of the distribution (see Figure \ref{fig:CVaRDef}).
% \footnote{By $\mbox{CVaR}_\delta[X] \le 0 \Rightarrow \mathbb{P}[X \geq 0] \le 1-\delta$, a bounded CVaR indicates  a probabilistic constraint. } %With a larger $\delta$, there's a smaller possibility for the violation of the probabilistic constrained $X  \le 0$.
%CVaR for a random variable $\zeta$ is the expectation of this random variable when its cumulative distribution function(CDF) $F(\zeta)\geq \delta$, \ie 
% \begin{align}\label{eq:CVaRtailE}
% \mbox{CVaR}_\delta[X]:=\E\left[X | X \geq F_X^{-1}(\delta) \right]. 
% \end{align}
% Here, $\E$ stands for expectation.

For power flows over a certain distribution line, $\mbox{CVaR}_\delta$  measures the average value among the top $1-\delta$ fraction of the largest power flow values. Thus, by constraining the $\mbox{CVaR}_{0.99}$ value of the line flow below a threshold $\ol{b}$, for example, implies that not only 99\% of the power flows are below $\ol{b}$, but also the average power flow among the worst (largest) 1\% is below $\ol{b}$. In other words, the $\mbox{CVaR}$-based constraint restricts both the probability and the extent of the line limit violation. Following \cite{rockafellar2002conditional}, CVaR  is defined as
\vspace{-0.2in}
\begin{align}
\mbox{CVaR}_\delta[X] :=\underset{t\in \Rset}{\text{minimize}}\left\{t+\frac{1}{1-\delta}\mathbb{E}[X-t]_+\right\}\label{eq:CVarProbHold}
\end{align}
% \vspace{-0.1in}
for random variables $X$ with general distributions.
% which is convex and simplifies the computation for CVaR\footnote{Eq.(\ref{eq:CVarProbHold}) provides a CVaR measure for discontinuous random variable $X$.}.
% \bose{Edited till here...}

Given the distribution of $ \v{p}_0$, the risk-aware access limit auction becomes,
%\begin{subequations}
%    \begin{align}
%    & \underset{\v{\ul{C}}, \v{\ol{C}}}{\text{maximize}} && \sum_{k=1}^K \varphi_k(\v{\ul{C}}_k,\v{\ol{C}}_k)-\mathbb{E}[J(\ol{\Pbf}(\omega),\ul{\Pbf}(\omega))], 
%     \\
%    & \text{such that} 
%    &&\v{\ol{C}}_k^{\mbox{\tiny min}} \le \v{\ol{C}}_k , ~~\v{\ul{C}}_k^{\mbox{\tiny min}} \le \v{\ul{C}}_k , \forall k
%    \\
%    &&&  \v{\ol{P}}(\omega) \le \v{\ol{P}}^{\mbox{\tiny max}}, ~~\v{\ul{P}}(\omega) \le \v{\ul{P}}^{\mbox{\tiny max}},
%    \\
%    &&& \v{\ol{P}}(\omega) = \sum_{k=1}^K \v{\ol{C}}_k + \v{p}_0(\omega) ,
%    \\
%    &&&
%    \v{\ul{P}}(\omega) = \sum_{k=1}^K \v{\ul{C}}_k - \v{p}_0(\omega) ,
%    \\ 
%    &&& 
%    \v{p} =\sum_{k=1}^K \v{p}_k,
%    \\
%    &&&
%    \v{\ul{b}}
%    \leq \mbox{CVaR}_\delta[\v{A}(\v{p} + \v{p}_0(\omega))] 
%    \leq 
%    \v{\ol{b}},
%    \\
%  %  &&& \v{p}_0(\omega) \sim f(\omega),
%%    \\
%    &&& \text{for all } \v{p}_k \in [-\v{\ul{C}}_k, \v{\ol{C}}_k],
%    \end{align}
%    \label{eq:auction.CVaR}
%\end{subequations}
\begin{subequations}
    \begin{align}
    & \underset{\v{\ul{C}}, \v{\ol{C}}}{\text{maximize}} && \sum_{k=1}^K \varphi_k(\v{\ul{C}}_k,\v{\ol{C}}_k)-\mathbb{E}[J(\ol{\Pbf},\ul{\Pbf})], 
     \\
    & \text{subject to} && \eqref{eq:ROC}-\eqref{eq:Rul}\nn,
    % \v{\ol{C}}_k^{\mbox{\tiny min}} \le \v{\ol{C}}_k , ~~\v{\ul{C}}_k^{\mbox{\tiny min}} \le \v{\ul{C}}_k , 
    \\
    % &&&  \v{\ol{P}} \le \v{\ol{P}}^{\mbox{\tiny max}}, ~~\v{\ul{P}} \le \v{\ul{P}}^{\mbox{\tiny max}},
    % \\
    % &~~~~~~~~\ol{\lambdabf}:&& \v{\ol{P}}= \sum_{k=1}^K \v{\ol{C}}_k + \v{p}_0,
    % \\
    % &~~~~~~~~\ul{\lambdabf}:&&
    % \v{\ul{P}} = \sum_{k=1}^K \v{\ul{C}}_k - \v{p}_0,
    % \\ 
%    &&& 
%    \v{p} =\sum_{k=1}^K \v{p}_k,
%    \\
    &&&
    \mbox{CVaR}_\delta[\v{A}(\sum_{k=1}^K \v{p}_k + \v{p}_0)] % \v{\ul{b}} \leq
    \leq 
    \v{\ol{b}}, \label{eq:CVaRPF}
    \\
    &&&
    \mbox{CVaR}_\delta[-\v{A}(\sum_{k=1}^K \v{p}_k + \v{p}_0)] 
    \leq -\v{\ul{b}}, \label{eq:CVaRPF2}
    \\
  %  &&& \v{p}_0(\omega) \sim f(\omega),
%    \\
    &&& \text{for all } \v{p}_k \in [-\v{\ul{C}}_k, \v{\ol{C}}_k], \text{for } k =1,...,K.
    \notag
    % \label{eq:RO-SCVaR}
    \end{align}
    \label{eq:auction.CVaR}
\end{subequations}
Compared to the robust network model in  \eqref{eq:auction}, the risk-constrained mechanism differs in three aspects. First, $\ol{\Pbf}, \ul{\Pbf}$ in this formulation is random, and hence, \eqref{eq:ROP} imposes the upper bounds for all possible values of $\v{p}_0$. Second, we consider the DSO's \emph{expected} operational cost to serve the random  $\ol{\Pbf}, \ul{\Pbf}$ in the objective function. In effect, we maximize the average induced social surplus. Third, and most importantly, we impose CVaR constraints on network limit violations in \eqref{eq:CVaRPF} and \eqref{eq:CVaRPF2} for all possible values of injections/withdrawals from assets controlled by DERAs within their acquired access limits. These constraints are such that for \emph{any} power transactions by DERAs within their access limits, they limit network constraint violation probabilities below $1-\delta$ and ensure that the average magnitude of those risky power flows/voltages remains within the specified limits. By imposing the constraint for all  $\v{p}_k \in [-\v{\ul{C}}_k, \v{\ol{C}}_k]$, this formulation inherits the benefits of decoupled DERA-DSO operations from the robust formulation. Specifically, any real-time DERA-ISO contract within the DERA's acquired limit imposes at most a pre-defined level of security risk to the distribution network. Said risk stems from the uncertainty in power transactions from DSO's customers alone for which they typically have statistics from historical data.

% enter into any contract with the ISO in real-time,  power transactions of DSO's customers or assets controlled by other DERAs. Rather, they must ensure that the assets they command operate within the access limits they acquire. DSO processes the uncertainty in power transactions of DSO's customers alone for which they typically have access to historical data.} 

Next, we present a scenario-approach to approximate \eqref{eq:auction.CVaR}. Consider $S$ independent and identically distributed samples $\v{p}_0[1], \ldots, \v{p}_0[S]$ for the injection of DSO's customers $\v{p}_0$. Using  \eqref{eq:CVarProbHold} and replacing all expectations with empirical means over $S$ samples, we arrive at the following optimization program for stochastic access allocation. The detailed derivation is relegated to Appendix \ref{sec:SO_Eq}.

\begin{subequations}
    \begin{align}
    & \underset{\substack{\v{\ol{C}}, \v{\ul{C}},\v{\ol{P}}[s], \v{\ul{P}}[s], \\
    \v{\ol{t}},  \v{\ul{t}},\v{\ul{\gamma}}[s], \v{\ol{\gamma}}[s]}}{\text{maximize}} && \sum_{k=1}^K\varphi_k(\v{\ul{C}}_k,\v{\ol{C}}_k)-\frac{1}{S}\sum_{s=1}^SJ(\ol{\Pbf}[s],\ul{\Pbf}[s]), \label{eq:SSSO}
    \\
    & \text{such that} 
    &&   \text{for } k = 1,\ldots,K, \ s= 1, \ldots, S,
    \nn \\
    &~~\ol{\etabf},\ul{\etabf}:&&\v{\ol{C}}_k^{\mbox{\tiny min}} \le \v{\ol{C}}_k , ~~\v{\ul{C}}_k^{\mbox{\tiny min}} \le \v{\ul{C}}_k , \label{eq:SO_capC}
    \\
    &~~\ol{\omegabf}[s],\ul{\omegabf}[s]:&&  \v{\ol{P}}[s] \le \v{\ol{P}}^{\mbox{\tiny max}}, ~~\v{\ul{P}}[s] \le \v{\ul{P}}^{\mbox{\tiny max}}, \label{eq: Pbd}
    \\
    &~~~~~~~~\ol{\lambdabf}[s]:&& \v{\ol{P}}[s] = \sum_{k=1}^K \v{\ol{C}}_k + \v{p}_0[s], 
    \\
    &~~~~~~~~\ul{\lambdabf}[s]:&&
    \v{\ul{P}}[s] = \sum_{k=1}^K \v{\ul{C}}_k - \v{p}_0[s],
    \\
    % & &&
    % \v{\ol{t}}+\frac{1}{(1-\delta)S}\sum_{s=1}^S \left( \v{A}_+ \v{\ol{P}}[s] \right. 
    % \nn
    % \\
    % &&&
    % \left. \quad + \v{A}_-\v{\ul{P}}[s]- \v{\ol{b}}-\v{\ol{t}}\right)_+ \le {\bf 0},\label{eq: SOol}
    % \\
&~~~~~~~~\ol{\betabf}[s]:&&   \v{A}_+ \v{\ol{P}}[s]+ \v{A}_-\v{\ul{P}}[s]- \v{\ol{b}}-\v{\ol{t}} \le \v{\ol{\gamma}}[s],\label{eq:SOepiol}
    \\
    &~~~~~~~~\ol{\alphabf}[s]:&&{\bf 0}\le \v{\ol{\gamma}}[s],
    \\    
    &~~~~~~~~\ol{\mubf}:&&
   \v{\ol{t}}+\frac{1}{1-\delta}\frac{1}{S}\sum_{s=1}^S\v{\ol{\gamma}}[s]\le {\bf 0},
    \\
    &~~~~~~~~\ul{\betabf}[s]:&&  \v{A}_-\v{\ol{P}}[s]+ \v{A}_+\v{\ul{P}}[s]+\v{\ul{b}}-\v{\ul{t}}\le \v{\ul{\gamma}}[s],
    \\
    &~~~~~~~~\ul{\alphabf}[s]:&&{\bf 0}\le \v{\ul{\gamma}}[s],
    \\  
    &~~~~~~~~\ul{\mubf}:&&
    \v{\ul{t}}+\frac{1}{1-\delta}\frac{1}{S}\sum_{s=1}^S\v{\ul{\gamma}}[s]\le {\bf 0}.\label{eq:SOepiul}    
    % \v{\ul{t}}+\frac{1}{(1-\delta)S} \sum_{s=1}^S\left(\v{A}_-\v{\ol{P}}[s] \right.
    % \nn
    % \\
    % &&&
    % \left. \quad + \v{A}_+ \v{\ul{P}}[s]+\v{\ul{b}}-\v{\ul{t}}\right)_+\le {\bf 0}.
    % \label{eq: SOul}
    \end{align}
    \label{eq:auction.3}
\end{subequations}
Sample average approximations for such stochastic programs are known to become more accurate with growing number of samples, as \cite{shapiro21lectures} suggests. Thus, the solution of \eqref{eq:auction.3} should approach that of \eqref{eq:auction.CVaR} as $S \to \infty$.
One expects that when $\delta \uparrow 1$, the stochastic model constraints \eqref{eq:SOepiol}-\eqref{eq:SOepiul} roughly approximate network security constraints \eqref{eq:ROol}-\eqref{eq:ROul} from the robust auction model. Per our experiments, even with $\delta=0.99$, the resulting social welfare can be significantly higher than from the robust auction ($\sim$20\%), upon tolerating only 1\% network security constraint violation risk at a few buses.

To define the settlement mechanism from the above auction, associate Lagrange multipliers with the constraints in \eqref{eq:auction.3} as shown and denote optimal values of variables with ($\star$). 
% where we consider $S$ independent and identically distributed samples $\{\v{p}_0^{(1)}, ..., \v{p}_0^{(s)}, ..., \v{p}_0^{(S)} \}$ for the injection of DSO's customers $\v{p}_0$, and the robustness constraint in (\ref{eq:RO-SCVaR}) is replaced by the injection limit scenarios $(\ul{\Pbf}_s,\ol{\Pbf}_s)$. See detailed derivations about (\ref{eq:auction.CVaR}) and (\ref{eq:auction.3}) in Appendix.

%Here, the equivalence of (\ref{eq:auction.CVaR}) and (\ref{eq:auction.3}) is proved in the appendix, with the property of CVaR and randomly samples $\{\v{p}_0^{(1)}, ..., \v{p}_0^{(s)}, ..., \v{p}_0^{(S)} \}$ for  $\v{p}_0(\omega)$. 

%Let $\omega \in \Omega$ denote the randomness of DSO's customers with $\Omega$ representing the field of uncertainties for renewable, load, etc.   We assume DSO knows the distribution for the net injection of DSO's customers $ \v{p}_0(\omega)$ which is a random parameter. Adopting the risk-based line security constraints, we formulate the stochastic access allocation by 

%Let $\Omega$ be the sample space of random net power injection $\pbf_0$ of DSO's customers and $\pbf_0(\omega), \omega \in \Omega$, a realization. We consider the following risk-constrained stochastic surplus maximization:

%To compute (\ref{eq:auction.CVaR}), we propose a scenario-based optimization below 

% \subsection{LMAP-S: Stochastic Locational Marginal Access Price}

\begin{definition}[LMAP-S]
For the stochastic allocation mechanism, the locational marginal access prices for injection and withdrawal access limits to the distribution network are defined by the vectors $ \ol{\Lambdabf}^\star:=\sum_{s=1}^S \ol{\lambdabf}^{\star}[s] \in \Rset^N$ and $\ul{\Lambdabf}^\star:=\sum_{s=1}^S \ul{\lambdabf}^{\star}[s] \in \Rset^N$, respectively, obtained from the optimal Lagrange multipliers $\ol{\lambdabf}^{\star}[s], \ul{\lambdabf}^{\star}[s]$ of \eqref{eq:auction.3}.
% the optimal dual solutions $\ol{\lambdabf}^{\star}_s \in \Rset^N, \ul{\lambdabf}^{\star}_s \in \Rset^N$ of (\ref{eq:auction.3}). Specifically, the injection price at bus $i$ is given by $ \ol{\rhobf}:=\sum_{s=1}^S \ol{\lambdabf}_s^{\star} $ (\$/MW) and the withdraw price by $\ul{\rhobf}:=\sum_{s=1}^S \ul{\lambdabf}_s^{\star}$ (\$/MW).  
%\sum_{i=1}^N( \ol{C}_k^{\star(i)}\sum_{s=1}^S\ol{\lambda}_s^{\star(i)}+\ul{C}_k^{\star(i)}\sum_{s=1}^S\ul{\lambda}_s^{\star(i)}).
\end{definition}
With LMAP-S, DERA $k$'s payment to the DSO is given by 
\beq
\label{eq:paySO}
\tilde{{\cal P}}_k(\v{\ol{C}}_k^{\star},\v{\ul{C}}_k^{\star} )=\ol{\Lambdabf}^{\star\intercal} \v{\ol{C}}_k^{\star} +\ul{\Lambdabf}^{\star\intercal} \v{\ul{C}}_k^{\star}.
\eeq
These locational prices are dependent on the forecast scenarios but are uniform at each distribution bus. DERA $k$'s induced surplus in the stochastic auction equals 
\begin{align}
    \tilde{\Pi}_k^{\mbox{\tiny DERA}}=\varphi_k(\v{\ul{C}}^{\star}_k,\v{\ol{C}}^{\star}_k) -\tilde{{\cal P}}_k(\v{\ol{C}}_k^{\star},\v{\ul{C}}_k^{\star} )
    \label{eq:Pi.DERA.def.2}
\end{align}
and the DSO's sample average surplus equals
\begin{align}
    \tilde{\Pi}^{\mbox{\tiny DSO}}
    &:= \sum_{k=1}^K {\cal P}_k(\v{\ol{C}}_k^{\star},\v{\ul{C}}_k^{\star} )\\
    &\quad -\frac{1}{S}\sum_{s=1}^S(J(\ol{\Pbf}^{\star}[s],\ul{\Pbf}^{\star}[s])-J(\v{\ol{p}}_0[s],\v{\ul{p}}_0[s])).\nn
\end{align}
With this notation, we now present the properties of the stochastic allocation in the next result.
% ; the proof is in Appendix \ref{sec:ProfProp4}). 

\begin{figure*}[!t]
   \vspace{-0.3in}
% \begin{strip} 
% \begin{align*}
\begin{equation}
\begin{array}{cc}
\ol{\Lambdabf}^\star
&=\sum_{s=1}^S \frac{1}{S}{\nabla_{\ol{\Pbf}[s]}J(\ol{\Pbf}^{\star}[s],\ul{\Pbf}^{\star}[s])} 
 + \sum_{s=1}^S \left( \Abf_+^{\T} \ol{\betabf}^{\star}[s]+\Abf_-^{\T} \ul{\betabf}^{\star}[s] + \ol{\omegabf}^{\star}[s] \right),
\\
\ul{\Lambdabf}^\star
&=\sum_{s=1}^S \frac{1}{S} {\nabla_{\ul{\Pbf}[s]}J(\ol{\Pbf}^{\star}[s]),\ul{\Pbf}^{\star}[s])} + \sum_{s=1}^S \left( \Abf_+^{\T} \ul{\betabf}^{\star}[s]+\Abf_-^{\T} \ol{\betabf}^{\star}[s] + \ul{\omegabf}^{\star}[s] \right).
\end{array} %\overline{\rule{3cm}{0.4pt}}
\label{eq:LMAP-seq}
\end{equation}
% \end{align*}
% \end{strip}
   % \vspace{-0.3in}
\hrule
\end{figure*}
%\vspace{-0.2in}

\begin{proposition}\label{Prop:LAP-SO}
The following statements hold for the stochastic network access allocation \eqref{eq:auction.3}:
\begin{enumerate}[label=(\alph*), leftmargin=*]
    \item LMAP-S satisfies \eqref{eq:LMAP-seq}.
    \item $\tilde{\Pi}^{\mbox{\tiny DSO}} \geq 0$. \item $\tilde{\Pi}_k^{\mbox{\tiny DERA}} \geq 0$, if $\varphi_k(\v{0},\v{0})\geq 0$ and one of the following two conditions holds: (a)  $\v{\ol{C}}_k^{\mbox{\tiny min}}={\bf 0}, \v{\ul{C}}_k^{\mbox{\tiny min}}={\bf 0}$, or (b)
the constraints in \eqref{eq:SO_capC} are non-binding at optimality, i.e.,  $\v{\ol{C}}_k^\star > \v{\ol{C}}_k^{\mbox{\tiny min}}$,$\v{\ul{C}}_k^\star >  \v{\ul{C}}_k^{\mbox{\tiny min}}$.
\item When $J$ is linear and homogeneous across buses ($J(\ol{\Pbf}[s],\ul{\Pbf}[s]) = \sum_{i=1}^N  \ol{J} \cdot \ol{P}^{(i)}[s] + \sum_{i=1}^N \ul{J} \cdot \ul{P}^{(i)}[s] $) and the constraints in \eqref{eq: Pbd} are non-binding at all buses and scenarios, then $ \ol{\Lambda}^{(n)\star}\geq \ol{\Lambda}^{(m)\star}$ and $\ul{\Lambda}^{(n)\star}  \geq \ul{\Lambda}^{(m)\star}$, if bus $n$ is an ancestor of bus $m$.
\end{enumerate}
\end{proposition}

Overall, this result shows that the stochastic auction outcome behaves similarly to the robust counterpart recorded in Propositions~\ref{Prop:LAP}-\ref{Prop:PM2}. Specifically, LMAP-S admits a sensitivity interpretation and is monotonic along the distribution feeder under similar sufficient conditions as LMAP-R. The resulting settlement
% We derive the property of LMAP-S in the first bullet in  Proposition \ref{Prop:LAP-SO} (proof provided in Appendix \ref{sec:ProfProp4}) running parallel to Proposition \ref{Prop:LAP} of LMAP-R. Both propositions reveal how the marginal bid-in cost of DSO and the binding network security constraint influence LMAP. The average influences over multiple scenarios for LMAP-S are illustrated by  \eqref{eq:LMAP-seq}.
covers DSO's operating cost (on average) and DERAs have nonnegative surpluses under similar sufficient conditions as the robust auction.

\section{An Illustrative Example}
\label{sec:example.4}

\begin{figure}[!htb]
   \centering
   \vspace{-0.1in}
    \includegraphics[scale=1.15]{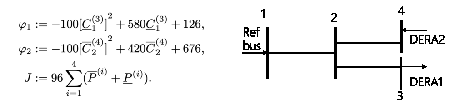}%ROtoy4busF
    \caption{A 4-bus distribution network example. Left: Bid-in functions of DERAs and DSO; right: network topology.}
    \label{fig:4busTEX}
% \end{wrapfigure} 
\end{figure}

We illustrate the properties of our access allocation mechanisms via a 4-bus network example with two different DERAs operating at two of the buses as shown in the right of Fig. \ref{fig:4busTEX}. The bids of the DERAs and the DSO's cost are shown in the left of Fig. \ref{fig:4busTEX}.
% \begin{align}
%     \begin{aligned}    
%     \varphi_1 &:=-100{[\ul{C}^{(3)}_1]}^2 + 580\ul{C}_1^{(3)} + 126,
%     \\
%     \varphi_2 &:=-100{[\ol{C}^{(4)}_2]}^2+420\ol{C}_2^{(4)} + 676,
%     \\
%     J &:= 96 \sum_{i=1}^4(\ol{P}^{(i)}+\ul{P}^{(i)}).
%     \end{aligned}
% \end{align}
Capacity limits for lines 2-3, 2-4, and 1-2 are taken as 1,1,2 (p.u.), respectively. The minimum access requirements of DERAs are zero, \ie $\v{\ol{C}}_k^{\mbox{\tiny min}}=\mathbf{0}, \v{\ul{C}}_k^{\mbox{\tiny min}}=\mathbf{0}$, for $k=1,2$, and the maximum access available is $\v{\ol{P}}^{\mbox{\tiny max}}=1$ p.u., $\v{\ul{P}}^{\mbox{\tiny max}}=1$ p.u. over all buses. DSO's customers have injections ranging over $[-0.15, 0.15]$ p.u. at all buses. We ignore voltage constraints for simplicity.
% , and the \tcr{LinDistFlow model is defined via $\Abf$, as shown in Fig.~\ref{fig:4busTEX}.}  
For the stochastic model, we set $\delta=0.9$ and draw 2000 scenarios  for $\v{p}_0$ for which each entry is i.i.d. truncated normal distributions with mean zero and standard deviation ($\sigma$) of $0.05$ over $[-0.15, 0.15]$, all in per units. 
 % \begin{wrapfigure}{l}{0.15\textwidth}

% \usepackage{booktabs}
% \usepackage{multirow}
\begin{table}[]
\caption{Access allocation result for the 4-bus example}
\label{table:alloc.4-bus}
\centering
\begin{tabular}{@{}cccccc@{}}
\toprule
\multirow{2}{*}{Allocation}                                  & \multirow{2}{*}{Parameter}       & \multicolumn{4}{c}{Bus $i$} \\ \cmidrule(l){3-6} 
                                                             &                                  & 1   & 2   & 3      & 4      \\ \midrule
\multirow{4}{*}{Robust}                                      & $\underline{C}_1^{(i)\star}$     & 0   & 0   & 0.85   & 0      \\
                                                             & $\underline{\lambda}^{(i)\star}$ & 96  & 96  & 410    & 96     \\
                                                             & $\overline{C}_2^{(i)\star}$      & 0   & 0   & 0      & 0.85   \\
                                                             & $\overline{\lambda}^{(i)\star}$  & 96  & 96  & 96     & 250    \\ \midrule
\multicolumn{1}{l}{\multirow{4}{*}{Stoch ($\delta = 0.9$)}} & $\underline{C}_1^{(i)\star}$       & 0   & 0   & 0.91   & 0      \\
\multicolumn{1}{l}{}                                         & $\underline{\Lambda}^{(i)\star}$            & 96  & 96  & 397.18 & 96     \\
\multicolumn{1}{l}{}                                         & $\overline{C}_2^{(i)\star}$        & 0   & 0   & 0      & 0.91   \\
\multicolumn{1}{l}{}                                         & $\overline{\Lambda}^{(i)\star}$  & 96  & 96  & 96     & 237.16 \\ \bottomrule
\end{tabular}
\end{table}

% \begin{table}[]\label{table:alloc.4-bus}
% \caption{Access allocation result for the 4-bus example}
% \centering
% \begin{tabular}{llllll}\hline\hline
% \centering
%  Entry & Case & Bus 1 & Bus 2 &  Bus 3 &  Bus 4 \\\hline 
% DERA$_1$ $\underline{C}$ (p.u.)& RO & 0 & 0 & 0.85 & 0   \\
% LMAP-R $\underline{\lambda}$ (\$/p.u.)& RO & 96 & 96 & 410 & 96   
% \\\hline
% DERA$_2$ $\overline{C}$ (p.u.)& RO & 0 & 0 & 0 & 0.85   
% \\ 
% LMAP-R $\overline{\lambda}$ (\$/p.u.)& RO & 96 & 96 & 96& 250 
% \\
% \hline 
% DERA$_1$ $\underline{C}$ (p.u.)& SO & 0 & 0 & 0.91 & 0   \\
% LMAP-S $\underline{\Lambda}$ (\$/p.u.)& SO & 96 & 96 & 397.18 & 96   \\\hline
% DERA$_2$ $\overline{C}$ (p.u.)& SO & 0 & 0 & 0 & 0.91   \\ 
% LMAP-S $\overline{\Lambda}$ (\$/p.u.)& SO & 96 & 96 & 96& 237.16 \\\hline\hline
% \end{tabular}
% \end{table}
%----------------------
% \begin{table}[]\label{table:SO4-bus}
% \caption{Stochastic access allocation result of 4-bus toy example}
% \centering
% \begin{tabular}{llllll}\hline\hline
% \centering
%  Entry &  Bus 1 & Bus 2 &  Bus 3 &  Bus 4 &  Surplus (\$)\\\hline 
% DERA1 $\underline{C}$ (MW) & 0 & 0 & 0.91 & 0 & 209.51 \\
% DSO  &-- & -- & --& -- & 275.28  \\
% LMAP-S $\underline{\lambda}$ (\$/MW) & 96 & 96 & 397.18 & 96 & -- \\\hline
% DERA2 $\overline{C}$ (MW) & 0 & 0 & 0 & 0.91 & 759.61 \\
% DSO  & -- & -- & -- & -- & 129.08 \\
% LMAP-S $\overline{\lambda}$ (\$/MW) & 96 & 96 & 96 & 237.16 & -- \\\hline\hline
% \end{tabular}
% \end{table}
\begin{table}[]
\caption{Surplus distribution for the 4-bus example}
\label{table:surplus.4-bus}
\centering
\begin{tabular}{llllll}\toprule
\centering
Allocation &  DERA$_1$ & DSO & DERA$_2$ & Social Surplus \\
\midrule
Robust  & 198.25 & 282.6  & 748.25 & 1229.1 \\ 
Stoch ($\delta = 0.90)$  & 209.51 & 404.36   & 759.61 & 1373.48 \\ \bottomrule
\end{tabular}
\end{table}

% \vspace{-0.05in}
Results from the robust and stochastic auction mechanisms are included in Tables \ref{table:alloc.4-bus} and \ref{table:surplus.4-bus}. As Table \ref{table:alloc.4-bus} reveals, LMAP-R and LMAP-S along the bus-sequences 1-2-3 and 1-2-4 increase, aligned with the conclusions of Propositions \ref{Prop:PM2} and \ref{Prop:LAP-SO}. The surpluses of the DERAs and the DSO in Table \ref{table:surplus.4-bus} are non-negative and hence, align with the results of Propositions \ref{Prop:RA} and \ref{Prop:LAP-SO}. Notice that compared to the robust allocation model, the stochastic allocation yields lower access prices, higher allocations to DERAs, and higher surpluses for the DERAs and the DSO. In effect, tolerating a small security risk yields less conservative allocations supported by lower prices and higher social surplus. 
In practice, one should set the risk tolerances based on exhaustive simulations for which observed real-time violations of security constraints are deemed acceptable.
% As will become evident in the next section, similar trends continue to hold in larger distribution networks. In fact, the difference between the two allocation models for larger systems is more pronounced as only a few buses typically carry the burden for constraint violations, while the rest of the system contributes to a substantial improvement in social welfare.

% \bose{The last bit is hard to understand.}
%\input{ToySOEX_v1}
\section{Case Study on a 141-Bus Network}\label{sec:CaseStudies}

We consider a 141-bus radial distribution network from \cite{khodr08EPSR141} with 4 DERAs that aggregate households at different buses with different levels of behind-the-meter distributed generation (BTM DG). Simulations were performed on a personal computer with Intel(R) Xeon(R) Gold 6230R CPU @ 2.10GHz and 256 GB RAM. We used YALMIP \cite{Yalmip} with Gurobi 10.0.0 \cite{Gurobi} on MATLAB R2021a to solve the optimization problems. The robust auction took $<$1s; the stochastic  model with 500 scenarios took $<$5min and with 1500 scenarios took $<$1hr.

% \begin{figure}[htbp]
%     \centering
%     \includegraphics[width=0.48\textwidth]{Figs/case141ini.eps}
%     \caption{A 141-bus radial distribution network from \cite{khodr08case141}.}
%     \label{fig:CASE141}
% \end{figure}
Network parameters for this system, including the resistance, reactance, topology, and base values, were adopted from \cite{khodr08EPSR141,MatpowerCase141}. The base value for voltage was 12.47kV and for power was 10MW. We assumed a fixed power factor of 0.98 lagging across all buses and set the line flow limits to be 20MW for all branches. The BTM DG  for households under DERAs 1, 2, 3, and 4 were set to $0.2$kW, $1.2$kW, $3.2$kW, and $4.2$kW, respectively. DERAs 1-3 aggregated resources from all buses and DERA 4 only aggregated over buses 118-134. The bid-in utility function $\varphi_k$ of DERA$_k$ is assumed to equal the sum of $\varphi_k^{(i)}$ for each bus $i$, where DERA$_k$ operates; $\varphi_k^{(i)}$'s are reported in Table \ref{table:benefit141}. The derivation of $\varphi_k^{(i)}$ is relegated to \cite{ChenAlahmedMountTong23DERA}; a short explanation is included in Appendix \ref{sec:paper2.141}. Minimal access limits for the DERAs were assumed uniform across all buses, values for which are in Table \ref{table:benefit141}.
% Each DERA aggregates 50 households on these buses. 
DSO's operational cost was assumed to be the sum of quadratics, $\frac{1}{2}bx^2+ax$ with  $a=\$0.009/\mbox{kWh}, b=\$0.0005/(\mbox{kWh})^2$ for both the injection and withdrawal access at all buses.

Power injection $p_0^{(i)}$ from the DSO's customer at each bus $i$ was sampled from independent truncated normal distributions with mean $\mu=5$kW, standard deviation $\sigma$, truncated to $[\ul{p}_0^{(i)}, \ol{p}_0^{(i)}] = [\mu-3\sigma, \mu+3\sigma]$. We used these intervals for the robust allocation model, but utilized 1500 random samples within said intervals for the stochastic allocation results. 

% \bose{Even though you are using part II to construct the bids, refer to the table with the explicit bids. Cong: Do you mean Table IV is not enough?}

% \usepackage{booktabs}
\begin{table}[]
\caption{Bid-in utility function and minimum network withdrawal/injection  limits for DERAs at each bus}
\label{table:benefit141}
\centering
\begin{tabular}{@{}cccc@{}}
\toprule
DERA $k$ & $\varphi^{(i)}_k$                                     & ${\ul{C}}^{\mbox{\tiny min}}_k$ (kW) & ${\ol{C}}^{\mbox{\tiny min}}_k$ (kW) \\ \midrule
1    & $- 0.1{[\ul{C}^{(i)}]}^2 + 2.8\ul{C}^{(i)} - 1.655$ & 4.1                                & 0                                  \\
2    & $- 0.1{[\ul{C}^{(i)}]}^2 + 1.8\ul{C}^{(i)} + 1.513$ & 0                                  & 0                                  \\
3    & $- 0.1{[\ol{C}^{(i)}]}^2 + 0.2\ol{C}^{(i)} + 7.393$ & 0                                  & 0                                  \\
4    & $- 0.1{[\ol{C}^{(i)}]}^2 + 1.2\ol{C}^{(i)} + 2.833$ & 0                                  & 0                                  \\ \bottomrule
\end{tabular}
\end{table}
\vspace{-0.2in}
\subsection{Running the Robust Access Allocation}
% Here we demonstrate the robust access allocation proposed in Sec.~\ref{sec:AccessRight}.  %. Thus the uncertain interval of $\v{p}_0$ had $\v{\ul{p}}_0=(-\mu+3\sigma){\bf 1}, \v{\ol{p}}_0=(\mu+3\sigma){\bf 1}$. 
% \bose{There is probably a better way to show the prices. The figure is becoming hard to read otherwise. Is this a MATLAB figure? I can help to change it. Cong: It is matlab figure.}

\begin{figure}[htbp]
    \centering
    \includegraphics[scale=0.6]{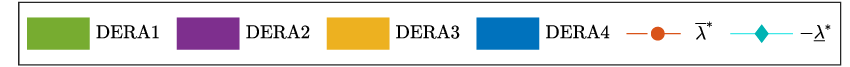} \includegraphics[width=0.45\textwidth]{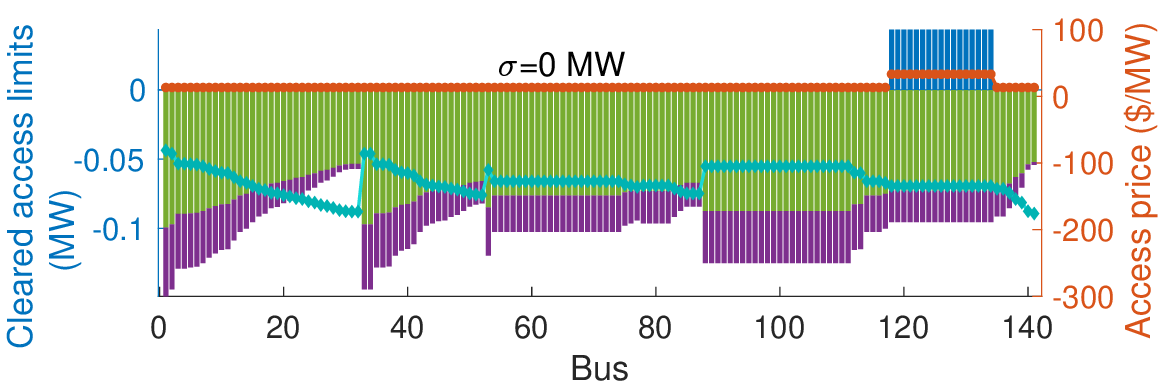}
    \includegraphics[width=0.45\textwidth]{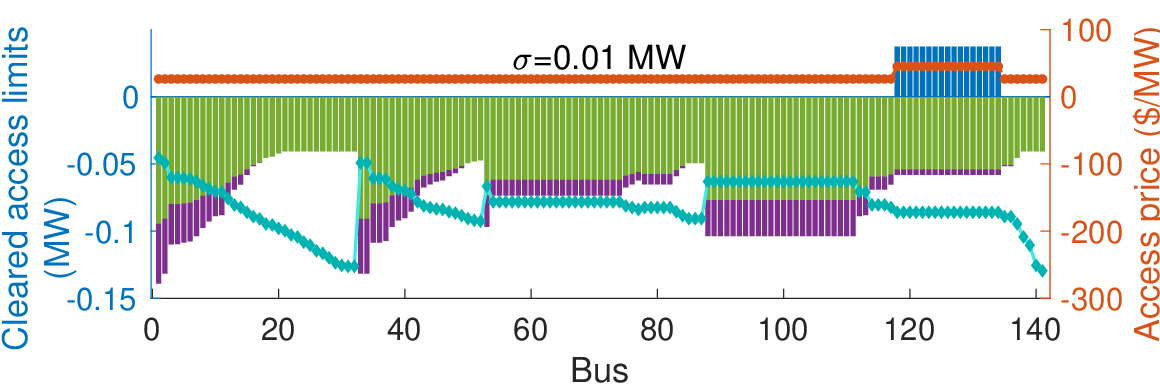}    %\includegraphics[scale=0.45]{Figs/MCR025p_0.eps}
    \caption{Auction results over the 141-bus distribution network. Left y-axis: cleared access limits, right y-axis: access clearing prices}
    \label{fig:MC}
\end{figure}

The access allocation results for the 4 DERAs under different $\sigma$ are illustrated in Fig. \ref{fig:MC}. The positive and negative segments of the left y-axis respectively represent the allocated injection and withdrawal ranges. The right y-axis shows the injection and withdrawal access prices. We show the injection access price over the positive segment of the right y-axis, and the negative range of the right y-axis shows the opposite number for the withdrawal access price. The plots reveal that cleared access limits for DERAs were smaller with higher $\sigma$. Such a trend is expected as higher $\sigma$ increases the burden from distribution utility's customers on the distribution network, implying a lesser share of the pie available to the DERAs. This burden makes network access more expensive. This manifested in higher locational access prices when  $\sigma$ was larger. 

% \vspace{-0.05in}

DERAs 1 and 2 had lower BTM DG compared to DERAs 3 and 4. Thus, DERAs 1 and 2 behaved as net consumers, who bid for and received withdrawal access, as shown in  Fig.~\ref{fig:MC}. DERAs 3 and 4, on the other hand, largely acted as net power producers, and they purchased injection access through the auction at buses they operated. We remark that DERA 3 has a small access limit across all buses with $\sigma=0$ MW, and it vanishes with $\sigma=0.01$ MW. DERA 3 commands less BTM DG than DERA 4, and the surplus it can generate for its customers is lower. As a result, it bids a higher induced utility, compared to DERA 4. Consequently, its cleared access remains lower than that of DERA 4. With the highest BTM DG, DERA 4 has the largest incentive to acquire injection access at the buses it operates at, i.e., buses 118 - 134.

% because of the higher DSO operation cost. This was because DSO robustly allocated most network accesses to its own customers implied by the large STD. So we observed that  DERA 3 purchased some injection accesses when $\sigma=0$ MW but no access when $\sigma=0.01$MW. And the accesses purchased by DERA 1 and DERA 2 also decreased when STD increased.

The access prices varied by location. As Proposition \ref{Prop:LAP} reveals, prices must be uniform unless either network constraints or maximum injection/withdrawal limits are binding. Indeed, with both values of $\sigma$, voltage constraints at buses 52 and 141 were binding, leading to locationally varying access prices. Moreover, the figures suggest that prices are monotonic only over certain segments of the distribution network. However, as the network structure in \cite{khodr08EPSR141} reveals, the price \emph{is} monotonic along paths away from the substation. Notice that price monotonicity is only guaranteed by Proposition \ref{Prop:PM2} with linear cost structures for the DSO. Our numerical results are derived with quadratic DSO costs, and yet, the conclusion of Proposition \ref{Prop:PM2} holds, implying that the conditions for the result as stated are only sufficient, but not necessary.

% Binding line and voltage limits caused access prices to vary with location. Access prices for  withdrawal $\ul{\lambda}_i^\star$ indeed were monotonic away from the distribution feeder, until reaching buses 32, 52, 87, 130, and 141. And  it's caused by  the binding voltage constraints at buses 52 and 141. This validated Proposition~\ref{Prop:PM2}. \bose{I do not understand what is being said here. Price monotonicity along the various paths away from the distribution feeder were broken at all these 5 locations just because of two voltage constraints? Cong: From Fig.5 and Fig,6 we can see the price monotonic trend is along different distribution feeders.}
%Locational allocation prices were gradually increasing along the branch until reaching the bus with  binding voltage constraints. 

% \subsubsection{Surplus Distribution}

% \begin{figure}[htbp]
%     \centering
%     %\includegraphics[scale=0.5]{Figs/CustomerSurplus4.eps}
%     \includegraphics[scale=0.5]{Figs/SurplusDistributionRO4.eps} %\includegraphics[scale=0.5]{Figs/DERAsurplusAll.eps}
%     \caption{Surplus distribution among DERAs, households, and DSO.}
%     \label{fig:SurplusProsumer}
% \end{figure}

\begin{table}[]
\caption{Variation of DERA surplus with $\sigma$}
\label{table:DERASS}
\centering
\begin{tabular}{@{}ccccc@{}}
\toprule
\multirow{2}{*}{DERA} & \multicolumn{4}{c}{$\sigma$ (MW)}       \\ \cmidrule(l){2-5} 
                      & 0       & 0.004   & 0.006   & 0.008   \\ \midrule
1                     & 599.54  & 488.00  & 431.20  & 369.41  \\
2                     & 324.07  & 291.43  & 277.58  & 265.01  \\
3                     & 1043.85 & 1042.54 & 1042.41 & 1042.41 \\
4                     & 80.18   & 76.74   & 75.09   & 73.49   \\ \bottomrule
\end{tabular}
\vspace{-0.1in}
\end{table}
% In this robust access allocation, surpluses of DERAs, households, and DSO are shown in Fig.~\ref{fig:SurplusProsumer} and Table~II, and the total height of the bar in Fig.~\ref{fig:SurplusProsumer} represents the social surplus. When STD $\sigma$ of the utility's customer power injection $\pbf_0$ increased,  the social surplus, DERA surplus, and DSO surplus decreased. This was because DSO robustly sold fewer network accesses when the STD $\sigma$ was large, as is shown in Fig.~\ref{fig:MC}. Thus the optimal energy scheduling of some DERAs cannot be realized due to the limited network injection/withdrawal accesses. The main part of the social welfare was the household surplus, which stayed the same  when STD changed because the DER aggregation method in the Part II paper maintained a constantly competitive surplus to households with customers' incumbent utility companies. 
The surpluses of the various DERAs are reported in Table \ref{table:DERASS}. 
% ~II, and the total height of the bar in Fig.~\ref{fig:SurplusProsumer} represents the social surplus. 
As evident, the surpluses reduce with higher $\sigma$. The larger the $\sigma$, the lower the DERAs' access to the network and consequently, their surpluses.

\vspace{-0.2in}
\subsection{Comparing the Robust and Stochastic Auctions}\label{sec:SOvsRO}%Results
\begin{figure}[htbp]
    \centering
    \vspace{-0.1in}
    \includegraphics[scale=0.4]{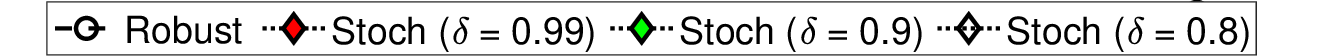} \includegraphics[scale=1.1]{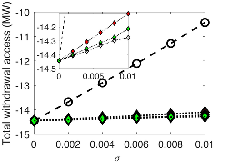}\includegraphics[scale=1.1]{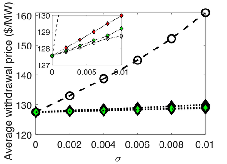}
    \includegraphics[scale=0.4]{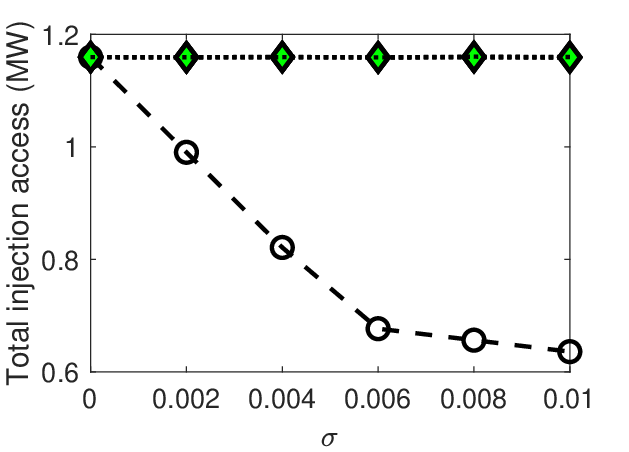}\includegraphics[scale=0.4]{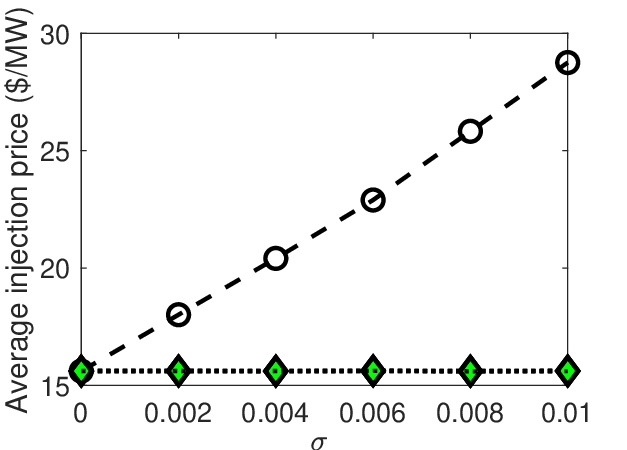}
    %\centering \includegraphics[scale=0.4]{Figs/Legend.eps}
    \caption{Top left: injection access, top right: injection price, bottom left: withdrawal access, bottom right: withdrawal price.}
    \label{fig:AccessPQ}
\end{figure}

%Assume the power injection profile from the utility's customer following Gaussian distribution, \ie  $p_0^i \sim {\cal N}(0, \sigma)$, $\forall i \in \{1,..., N\}$, 
For the stochastic access allocation models, we considered three different risk levels---$\delta=0.99,0.9,0.8$. 
% We considered three cases to compare the access allocation results under the robust and stochastic setting. The robust access allocation was simulated in Case 1 (abbreviation RO-3$\sigma$), and it had the same parameter setting as the previous section.   In Case 2 (abbreviation SO-$\delta0.99$), Case 3 (abbreviation SO-$\delta0.9$), and Case 4 (abbreviation SO-$\delta0.8$), the stochastic access allocation with constrained CVaR was considered, and $\delta=0.99, 0.9, 0.8$ respectively. Still, the power injection profile $\v{p}_0$ from the utility's customer at each bus followed a truncated normal distribution with mean $\mu=5$kW,  STD $\sigma$ and the truncated interval $p_0^i\in [\mu-3\sigma, \mu+3\sigma], \forall i$. The robust allocation only considered the truncated interval for the worst case, while the stochastic allocation mechanism  sampled 1500 scenarios to compute the SO-CVaR model.
% \subsubsection{Access Clearing Price and Quantities}
The allocation results for the withdrawal access and price are shown at the top of Fig.~\ref{fig:AccessPQ}.  Negative values indicate withdrawal access and positive values encode  injection access. Compared to the robust allocation mechanism, the stochastic model had larger withdrawal  accesses cleared and lower prices for those access limits. The difference in the outcomes from the robust and the stochastic models are substantial, even with a high risk parameter of $\delta=0.99$. With more uncertainty (larger $\sigma$), the cleared  withdrawal  accesses were lesser and allocation prices were higher---a trend we already observed with the robust allocation model.

% The allocation results for the injection access and price are shown at the bottom  of Fig.~\ref{fig:AccessPQ}. When the system had more uncertainty (larger $\sigma$), the stochastic allocation mechanism in Case 2-4  only saw slight differences since there was no scarcity caused by binding network operation constraints when allocating the injection access. However, the robust allocation mechanism  witnessed an obvious decrease in the cleared  injection accesses  and an increase in the prices. 

\vspace{-0.05in}

% \bose{It is important to talk more precisely about the sampling method here. Did you add independent Guassians to each bus? In that case, you have very independent demand variations, where in reality, the variations in demand are highly correlated. In fact, that correlation will majorly benefit the stochastic allocation. Let's talk about this.}

% \begin{figure}[htbp]
%     \centering
%     \includegraphics[scale=1.3]{Figs/SocialWelfare1500F.eps}
% \caption{Social surplus. }
%     \label{fig:SS}
% \end{figure}
% \subsubsection{Social surplus}
% Social surpluses of different cases are illustrated in Fig.~\ref{fig:SS}\footnote{The social surplus here is the optimal value of (\ref{eq:SSRO}) and  (\ref{eq:SSSO}). But the social surplus in Fig.~\ref{fig:SurplusProsumer} is the customer surplus plus the optimal value of  (\ref{eq:SSRO}).}. In stochastic allocation Case 2 (SO-$\delta0.99$), by tolerating the 1\% voltage violation risk, the social surplus increased by 20\% at $\sigma=0.01$MW, compared to the conservative robust allocation mechanism in Case 1 (RO-3$\sigma$). This shows that with risk tolerance $1\%$ for the network security constraints, the stochastic allocation significantly increased the  social surplus compared to the over-conservative robust setting. 

% \subsubsection{ Revenue Adequacy of DSO and Individual Rationality of DERA}

%----------DSO stuff is not required.

\begin{figure}[htbp]
    \centering
     \includegraphics[scale=0.4]{Figs/Legend.eps}  
     \includegraphics[scale=1.1]{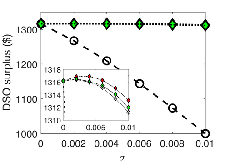}
    \includegraphics[scale=1.1]{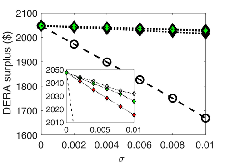}
    \includegraphics[scale=1.1]{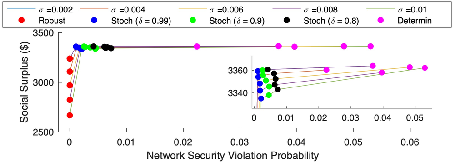}
\caption{Top left: DSO surplus. Top right: DERA surplus. Bottom: The network security violation probability and social surplus.}
    \label{fig:Surplusall}
\end{figure}

The DSO's surplus on the top left of Fig. \ref{fig:Surplusall} was nonnegative for all experiments. The same holds for the DERAs' surpluses, the aggregate value for which is plotted on the top right  of Fig. \ref{fig:Surplusall}. These verify Propositions \ref{Prop:RA}-\ref{Prop:PM2} and \ref{Prop:LAP-SO}.  
% .Such a metric was nonzero in all cases.  This validated the nonnegative surplus of DSO.  Additionally, the surplus of   DERA is shown on the right  of Fig.~\ref{fig:Surplusall}.  All cases had nonzero DERA surplus. Thus the above observations validated Proposition~\ref{Prop:RA} and Proposition~\ref{Prop:RACVaR}.  
The surpluses are higher under the stochastic model, compared to the robust counterpart. Correspondingly, the social surplus (the sum of DERAs' and DSO's surplus) at the bottom of Fig. \ref{fig:Surplusall}, is higher in the stochastic model.  In fact, the conservative robust allocation had DSO and DERA surpluses around 20\% lower than in the stochastic setting when $\sigma=0.01$MW. This observation suggests that even a small $\sim$1\% risk tolerance can drastically improve the surpluses of the auction participants. 

The variation of DERA surpluses with $\sigma$ is expected---largely, the DERAs' surpluses reduced with higher $\sigma$ that burdens the distribution network with higher possible injections from the DSO's customers. As a result, DERAs got lesser access limits with lower surpluses. %The influence of  to the robust optimization is obvious from the truncated interval $[\ul{p}_0^{(i)}, \ol{p}_0^{(i)}] $ in the parameter setting. When DSO's customer has larger $\sigma$, the equation (4d)-(4g) provide a smaller feasible domain for $\overline{\Cbf}_k$ and $\underline{\Cbf}_k$. Thus the optimal value of DERA surplus is smaller. Similarly, for the  stochastic model with larger $\sigma$, the equation (13d)(13e) has a larger possibility to sample a large $\pbf_0[s]$. This makes the feasible domain for $\overline{\Cbf}_k$ and $\underline{\Cbf}_k$ smaller. Therefore, the  DERA surplus is smaller when  $\sigma$ is larger.

%The DERA surplus decreases when $\sigma$ increases for both the robust and stochastic model. When  $\sigma$ is larger, DSO needs to partition more network access for DSO's customers, which have more volatile net injections, in order to obey the network voltage limit and line capacity limit. 

% As for the influence of randomness, it's observed that both DERA and DSO surplus decreased  when $\sigma$ increased. Such trends of DSO surplus and DERA surplus versus $\sigma$  were also aligned with the social surplus in Fig.~\ref{fig:SS}.  
%Additionally,  the more conservative robust allocation mechanism had less surplus for both DERA and DSO than that under the stochastic allocation mechanism with constrained CVaR. When the system had less uncertainty and thus smaller $\sigma$, both DERA and DSO  had higher  surplus.
\vspace{-0.05in}
% \tcr{figure text size >}

% \bose{I think you can remove Figure 7 altogether and have Figure 9 take its place. Talk about individual surpluses and then mention that the total surplus (which is the addition of all of them) shows a similar trend. Cong: I can remove Fig.7. How about keep Table V, which gives details about individual DERAs' surplus.}

% One might anticipate that network constraint violations might increase substantially under the risk-sensitive auction design compared to the robust version. From the bottom of Fig. \ref{fig:Surplusall}, we reiterate that violations in the forward auctions are anticipatory, and should not be conflated with real-time violations. In our numerical experiment, upon tallying across 141 buses and 1500 scenarios, there were no upper and only 0.0295\% lower voltage constraint violations, and that too, largely concentrated at buses 52 and 141. At bus 141, the average violations for the lower bounds on voltages were 0.4\%. Thus, allowing small constraint violations in our CVaR-based  auction amounted to small probabilities and extents of violations compared to the robust design.

We also compared our results with a \emph{deterministic} model that computed the access against the {average} of all sampled scenarios by running \eqref{eq:auction.3} with that average scenario. As such an auction model ignores the uncertainty in the power injection/withdrawal scenarios, network security constraint violation probabilities were 7-30 times higher than that of the stochastic auction, as  Fig. \ref{fig:Surplusall} reveals. Recall that uncertainties in operating conditions result from two sources--the first is the natural variation in the real-time DERA-ISO interactions over time, and the second is the aleatory uncertainty in possible real-time system conditions as visible at the forward auction stage. Undoubtedly, any uncertainty model that a DSO adopts, will affect the outcomes of our auctions. From practical use, a DSO must calibrate the uncertainty model, the resulting auction outcomes, and its implications for real-time DERA-ISO transactions through exhaustive simulations.

\section{Conclusions}\label{sec:Conclusion}

We have proposed a DSO-DERA-ISO coordination mechanism for multi-DERA participation in the wholesale market. Through a forward auction and welfare-maximizing robust and risk-sensitive market clearings, the proposed mechanism allocates network access limits to DERAs based on their willingness to pay for the  access of the DSO-operated network. The key advantage of our proposed coordination mechanism is that it decouples DSO-DERA-ISO operations--the DSO-DERA forward auction in our model computes  operating envelopes for real-time DERA-ISO transactions. This salient feature makes the proposed solution compatible with existing jurisdictional boundaries of DSOs and ISOs and yet, guarantees distribution network security with minimal coorindation.

Several relevant issues remain outside the scope of this work, requiring further investigation. We have not considered possible topology changes of the distribution network in our auction designs. We anticipate that a security-constrained  version of our auction can be designed similarly to the security-constrained unit commitment and  economic dispatch problem for wholesale market clearing. In this work, we have not addressed the question of bid/offer formation for DERAs. The overall market efficiency of our designs rests on this bid/offer formation process with price-taking and price-making DERAs. See our parallel effort in \cite{ChenAlahmedMountTong23DERA} for a step towards such an analysis.
We have not delved into the details of the DSO's bid-in operational costs. Defining the scientific basis for a regulatory framework to calculate such costs, building on insights from \cite{Yeddanapudi08Maintainance} remains an interesting direction for future research. Admittedly, our simulations are limited in scope, designed primarily to study the properties of our design. As we have repeatedly pointed out, more realistic empirical analyses on larger systems are required to validate the practical efficacy of our designs.

\begin{figure*}[tbp]
    \centering
\begin{minipage}{.4\textwidth}
    \centering
   \includegraphics[width=0.75\textwidth]{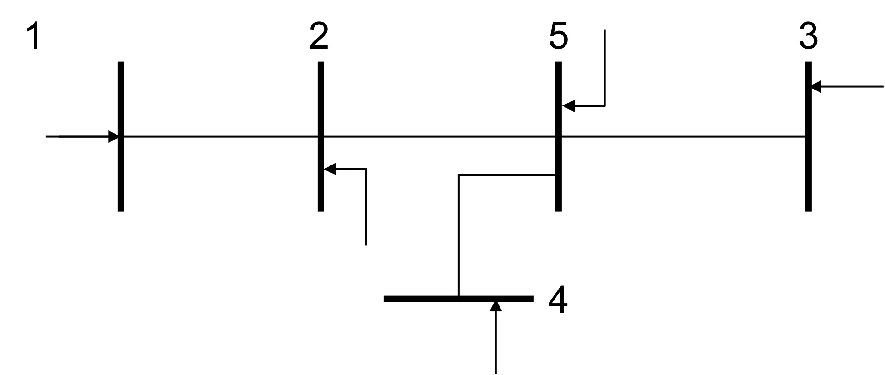}
\end{minipage}%
 \begin{minipage}{.6\textwidth}
    {\small
\begin{align*}
\Lbf =\begin{blockarray}{cccccc}
1 & 2 & 3 & 4 & 5 \\
\begin{block}{(ccccc)c}
-1&1&0&0&0 \\
0&-1&0&0&1 \\
0&0&1&0&-1 \\
0&0&0&1&-1 \\
\end{block}
\end{blockarray}
; \quad \Sbf=
\begin{blockarray}{cccccc}
1 & 2 & 3 & 4 & 5 & \text{Bus/Line}\\
\begin{block}{(ccccc)c}
0&1&1&1&1 & 2-1 \\
0&0&1&1&1 & 5-2\\
0&0&1&0&0 & 3-5\\
0&0&0&1&0 & 4-5\\
\end{block}
\end{blockarray}
\end{align*}
}
\end{minipage}   
\caption{A 5-bus radial distribution network example and its associated matrices.}
\label{fig:c}
\end{figure*}
\vspace{-0.1in}

{
\bibliographystyle{IEEEtran}
\bibliography{BIB}
}

%\newpage
\section{Appendix}
\label{sec:Appendix}
 \subsection{The Linearized Power Flow Model}\label{sec:MLPF}

\newcommand{\pinj}{\v{p}}
\newcommand{\fP}{f_{\textrm{P}}}
\newcommand{\fQ}{f_{\textrm{Q}}}
\newcommand{\fPbf}{\v{f}_{\textrm{P}}}
\newcommand{\fQbf}{\v{f}_{\textrm{Q}}}

The power flow model we use resembles the linearized distribution flow model described in \cite{Low19PowerSystemAnalysis} that is based on \cite{baranWu89TPDdistFlow}.
Consider a radial distribution network with $N$ buses in the node set $\Nc$ and $N-1$ branches in the edge set $\Ec$, represented by a directed graph $\Gc=(\Nc,\Ec)$. Call bus 1 the reference bus, and assign the edge directions arbitrarily. Let $\pinj \in \Rset^{N}$ be the vector of real power injections in per unit (p.u.) across the network. With a fixed power factor, the vector of net reactive power injection in p.u. then becomes $\alpha \pinj$, and the linearized power flow equations are described by
% \footnote{\tcb{ We adopt the constant power factor assumption to  simplify the model and proof in this paper. All theoretical results can be extended to the case without the constant power factor assumption.}}
\begin{subequations}
\begin{align}
\sum_{k:j\to k} \fP^{(j,k)}
&=\sum_{i:i\to j} \fP^{(i,j)}+p^{(j)},\label{eq:LinDist1}
\\
\sum_{k:j\to k} \fQ^{(j,k)}
&=\sum_{i:i\to j} \fQ^{(i,j)}+\alpha p^{(j)},\label{eq:LinDist2}
\\
{[v^{(j)}]}^2-{[v^{(k)}]}^2
&=2 r^{(j,k)} \fP^{(j,k)} + 2 x^{(j,k)} \fQ^{(j,k)}\label{eq:LinDist3}
% \end{cases} 
\end{align}
\label{eq:LinDistFlow}
\end{subequations}
for all $j \in \Nc$ and $j \to k \in \Ec$. Here, we use the notation $j\to k$ to denote an edge from $j$ to $k$. Also, voltage magnitudes in p.u. across the network are collected in $\v{v} \in \Rset^{N-1}$, and the directed real and reactive power flows over the distribution lines are collected in $\fPbf \in \Rset^{N-1}$ and $\fQbf \in \Rset^{N-1}$, respectively. The relations in \eqref{eq:LinDistFlow} capture real and reactive power balances at all distribution buses, and the voltage magnitude variations across the distribution lines. The line between buses $j$ and $k$ are characterized by the resistance and reactance, $r^{(j,k)}$ and $x^{(j,k)}$, respectively. Denote their collections across the network by $\rbf \in \Rset^{N-1}, \xbf\in \Rset^{N-1}$. Next, we write \eqref{eq:LinDistFlow} more compactly using $\Gcal$-dependent matrices that we introduce next.

Define the \emph{incidence matrix} $\Lbf \in \Rset^{(N-1)\times N}$ of $\Gcal$ via
\beq
\label{eq:IM}
L^{(j,n)}=\begin{cases}
1,& \text{if } j=n \to k ~\text{for some} ~k,\\
-1,& \text{if } j=k \to n ~\text{for some} k,\\
0, &\text{otherwise}.\end{cases} 
\eeq
Next, define the \emph{path matrix} $\Sbf \in \Rset^{(N-1) \times N}$ of $\Gcal$ via
\beq
\label{eq:PM}
S^{(n,j)}=
\begin{cases}
1,& \text{if } j \in \text{undirected path between buses }1, n,\\
0, &\text{otherwise}.\end{cases} 
\eeq
For any bus $n$, the undirected path from bus $n$ to the reference bus 1 is unique, and hence, the condition in the above relation is well-defined. These definitions are illustrated through an example in Fig.~\ref{fig:c}.
Let $\tilde{\Lbf}, \tilde{\Sbf} \in \Rset^{(N-1)\times(N-1)}$ denote the \emph{reduced} incidence and path matrices obtained by eliminating the first columns (corresponding to the reference bus) of $\Lbf$ and $\Sbf$, respectively. Let $\tilde{\pbf} \in \Rset^{N-1}$ be the vector of real power injections $ \pbf$ without the column for the reference bus. With this notation, \eqref{eq:LinDist1} and \eqref{eq:LinDist2} can be summarized in 
\begin{align}
    \tilde{\Sbf} \tilde{\v{p}} = \fPbf,
\end{align}
with $\fQbf = \alpha \fPbf$, and  \eqref{eq:LinDist3} can be written as
\beq\label{eq:LinDistFlowMMInc}
% \begin{cases}
% \tilde{\Sbf}\tilde{ \pbf} =\fPbf,
% \\
2 \Dbf \tilde{\Sbf} 
\tilde{ \pbf} 
= \tilde{\Lbf} \tilde{\ubf}-[u^{(1)};{\bf 0}]^\intercal,
% \end{cases} 
\eeq
where we use the additional notation $\Dbf:=\diag(\rbf+\alpha \xbf)$, $\ubf \in \Rset^{N}$ defined by $u^{(i)}:={[v^{(i)}]}^2$ for $i \in \Nc$, and $\tilde{\ubf} \in \Rset^{N-1}$ as the reduced form of $\ubf$ with the first row removed (again, corresponding to the reference bus). The notation $[u^{(1)}; \bf{0}]$ is a row vector of zeros of appropriate dimension with $u^{(1)}$ augmented as the first element. 
Per \cite[p. 15]{bapat10graphs}, $\tilde{\Lbf}^{-1}=\tilde{\Sbf}^\intercal$, and both $\tilde{\Lbf}$ and $\tilde{\Sbf}$ are invertible. Thus, (\ref{eq:LinDistFlowMMInc}) admits the compact representation, as in \cite[p. 206]{Low19PowerSystemAnalysis},
\beq\label{eq:LinDistFlowMM}
\tilde{\Sbf}\tilde{ \pbf} =\fPbf,
\quad
2 \tilde{\Sbf}^\intercal \Dbf\tilde{\Sbf} 
\tilde{ \pbf}
=\tilde{\ubf}- \tilde{\Sbf}(1,:)^\intercal u^{(1)}.
\eeq 
Here, $\tilde{\Sbf}(1,:)$ denotes the first row of $\tilde{\Sbf}$. 

To arrive at the constraints, notice that voltages at all buses, save the reference bus, are typically constrained within a narrow band, $0.95^2 \mathbf{1} \leq \tilde{\ubf} \leq 1.05^2 \mathbf{1}$, equivalently,
\begin{align}
    \underbrace{0.95^2 \mathbf{1} - \tilde{\Sbf}(1,:) u^{(1)}}_{:=\ul{\ubf}}
    \leq 
    2 \tilde{\Sbf}^\intercal \Dbf\tilde{\Sbf} 
\tilde{ \pbf} 
    \leq 
    \underbrace{1.05^2 \mathbf{1} - \tilde{\Sbf}(1,:) u^{(1)}}_{:=\ol{\ubf}}.
\end{align}
Distribution line flows are constrained as
$${[\fP^{(j,k)}]}^2 + {[\fQ^{(j,k)}]}^2 \leq {[\ol{f}_\alpha^{(j,k)}]}^2$$
for each $(j,k) \in \Ec$ for some line limit $\ol{f}_\alpha^{(j,k)} > 0$. Call the collection of these limits $\ol{\fbf}_\alpha \in \Rset^{N-1}$. Using the constant power factor assumption, the line flow constraints are
\begin{align}
    \underbrace{-{\ol{\fbf}_\alpha}{({1+\alpha^2})^{-1/2}}}_{:=\ul{\fbf}}
    \leq 
    \fPbf = \tilde{\Sbf}\tilde{\pbf}
    \leq
    \underbrace{{\ol{\fbf}_\alpha}{({1+\alpha^2})^{-1/2}}}_{:=\ol{\fbf}}.
\end{align}
We then obtain the constraints in \eqref{eq:volLBUB} with\footnote{From \eqref{eq:PM}, it follows that $\tilde{\Sbf} \geq 0$ element-wise. Since $\v{D} \geq 0$ element-wise, we have $\Abf=\Abf_+$ and $\Abf_-=\mathbf{0}$ for our LinDistFlow model. } 
\begin{align}
    \Abf 
    := \begin{pmatrix} 
    \mathbf{0} & \tilde{\Sbf} 
    \\
    \mathbf{0} & 2 \tilde{\Sbf}^\intercal \v{D} \tilde{\Sbf}
    \end{pmatrix},
    \;
    \v{\ol{b}} 
    := \begin{pmatrix}
    \ol{\fbf} \\ \ol{\ubf}
    \end{pmatrix},
    \;
    \v{\ul{b}} 
    := \begin{pmatrix}
    \ul{\fbf} \\ \ul{\ubf}
    \end{pmatrix}.
    \label{eq:Abb.def}
\end{align}

When voltage constraints are ignored (e.g., in Sec.~\ref{sec:example.4}), the bottom row in \eqref{eq:Abb.def} is removed.

\vspace{-0.2in}
\subsection{Proof of Lemma~\ref{lemma:EqRO}}
Notice that $ \v{p}_k \in [-\v{\ul{C}}_k, \v{\ol{C}}_k]$, $\v{p}_0 \in [-\v{\ul{p}}_0, \v{\ol{p}}_0]$ implies
\begin{align}
    \v{p} := \sum_{k=1}^K \v{p}_k+ \v{p}_0 \in [-\ul{\v{P}}, \ol{\v{P}} ].
\end{align}
For an arbitrary vector $\v{a} \in \Rset^N$, we have
\begin{align}
    \max_{\v{p} \in [-\ul{\v{P}}, \ol{\v{P}} ]} \v{a}^\intercal \v{p} 
    &= \v{a}_+^\intercal \v{\ol{P}} 
    + \v{a}_-^\intercal \v{\ul{P}},
    \\
    \min_{\v{p} \in [-\ul{\v{P}}, \ol{\v{P}} ]} \v{a}^\intercal \v{p} 
    &= -\v{a}_+^\intercal \v{\ul{P}} 
    - \v{a}_-^\intercal \v{\ol{P}}.    
\end{align}
Thus, for some scalars $\ul{b} \leq \ol{b}$, imposing $\ul{b} \leq \v{a}^\intercal \v{p} \leq \ol{b}$ for all $\v{p} \in [-\ul{\v{P}}, \ol{\v{P}} ]$ is equivalent to 
\begin{align}
    \v{a}_+^\intercal \v{\ol{P}} 
    + \v{a}_-^\intercal \v{\ul{P}} \leq \ol{b},
    \quad
    -\v{a}_+^\intercal \v{\ul{P}} 
    - \v{a}_-^\intercal \v{\ol{P}} \geq \ul{b}.
\end{align}
Applying the above to all rows of $\v{A}$ completes the proof.

\vspace{-0.1in}
\subsection{Proof of Proposition \ref{Prop:LAP}}
\label{sec:prop1}
The Lagrangian function of \eqref{eq:auction.2} is
\begin{align}
\begin{aligned}
    & {\cal L}(\v{\ul{C}}, \v{\ol{C}}, \v{\ol{P}}, \v{\ul{P}}, \ol{\lambdabf}, \ul{\lambdabf}, \ol{\mubf}, \ul{\mubf})
    = J(\ol{\Pbf},\ul{\Pbf})-\sum_{k=1}^K\varphi_k(\v{\ul{C}}_k,\v{\ol{C}}_k) 
    \\
    &~~  +\ol{\lambdabf}^\T \left(  \sum_{k=1}^K \v{\ol{C}}_k + \v{\ol{p}}_0 -\v{\ol{P}} \right) +\ol{\etabf}^\T(\v{\ol{C}}_k^{\mbox{\tiny min}}-\v{\ol{C}}_k)
    \\
    &~~+\ul{\etabf}^\T(\v{\ul{C}}_k^{\mbox{\tiny min}}-\v{\ul{C}}_k) 
    +\ul{\lambdabf}^\T \left( \sum_{k=1}^K \v{\ul{C}}_k + \v{\ul{p}}_0 -\v{\ul{P}} \right)
    \\
    &~~    +\ol{\mubf}^\T(\v{A}_+ \v{\ol{P}} + \v{A}_-  \v{\ul{P}} -  \v{\ol{b}})
    +\ul{\mubf}^\T(\v{A}_+  \v{\ul{P}} + \v{A}_- \v{\ol{P}}+ \v{\ul{b}})
    \\
    &~~ +\ul{\omegabf}^\T( \v{\ul{P}}-\v{\ul{P}}^{\mbox{\tiny max}})+\ol{\omegabf}^\T( \v{\ol{P}}-\v{\ol{P}}^{\mbox{\tiny max}}).
    \end{aligned}
\end{align}
The necessary and sufficient Karush-Kuhn-Tucker optimality conditions for \eqref{eq:auction.2} then yield $\nabla_{\ol{\Pbf}}{\cal L}^\star = \nabla_{\ol{\Pbf}}{\cal L}^\star = 0$, where
\begin{align}\label{eq:KKTeq}
\begin{aligned}
\nabla_{\ol{\Pbf}}{\cal L}^\star &=\nabla_{\ol{\Pbf}}J(\ol{\Pbf}^\star,\ul{\Pbf}^\star)-\ol{\lambdabf}^\star+\Abf_+^{\T}\ol{\mubf}^\star + \Abf_-^{\T}\ul{\mubf}^\star+\ol{\omegabf}^\star,
\\
\nabla_{\ul{\Pbf}}{\cal L}^\star &=\nabla_{\ul{\Pbf}}J(\ol{\Pbf}^\star,\ul{\Pbf}^\star)-\ul{\lambdabf}^\star+\Abf_-^{\T}\ol{\mubf}^\star+\Abf_+^{\T}\ul{\mubf}^\star+\ul{\omegabf}^\star,
\end{aligned}
\end{align}
completing the proof.

\vspace{-0.2in}
\subsection{Proof of Proposition \ref{Prop:RA}}\label{sec:ProfProp2}
$\bullet$ Part (i): The KKT optimality conditions for \eqref{eq:auction.2} give
\begin{align}
    \begin{aligned}
    &\ol{\lambdabf}^{\star\intercal} \sum_{k=1}^K \v{\ol{C}}_k^{\star} +\ul{\lambdabf}^{\star\intercal} \sum_{k=1}^K \v{\ul{C}}_k^{\star} 
    \\
    &= (\ol{\Pbf}^\star-\v{\ol{p}}_0)^\intercal \ol{\lambdabf}^{\star}
    + (\ul{\Pbf}^\star-\v{\ul{p}}_0)^\intercal \ul{\lambdabf}^{\star} 
    \\
    &\overset{(a)}{=} (\ol{\Pbf}^\star-\v{\ol{p}}_0)^\intercal  \left( 
\nabla_{\ol{\Pbf}}J(\ol{\Pbf}^\star,\ul{\Pbf}^\star)+\Abf_+^{\T}\ol{\mubf}^\star + \Abf_-^{\T}\ul{\mubf}^\star+\ol{\omegabf}^\star    
    \right)
    \\
    & \; + (\ul{\Pbf}^\star-\v{\ul{p}}_0)^\intercal \left(
    \nabla_{\ul{\Pbf}}J(\ol{\Pbf}^\star,\ul{\Pbf}^\star)+\Abf_-^{\T}\ol{\mubf}^\star+\Abf_+^{\T}\ul{\mubf}^\star+\ul{\omegabf}^\star
    \right)
    \\
    &\overset{(b)}{\geq} (\ol{\Pbf}^\star-\v{\ol{p}}_0)^\intercal   
\nabla_{\ol{\Pbf}}J(\ol{\Pbf}^\star,\ul{\Pbf}^\star)  
    % \\
    % & \; 
    + (\ul{\Pbf}^\star-\v{\ul{p}}_0)^\intercal 
    \nabla_{\ul{\Pbf}}J(\ol{\Pbf}^\star,\ul{\Pbf}^\star)
    \\
     % &\geq (\ol{\Pbf}^\star-\v{\ol{p}}_0)^\intercal\nabla_{\ol{\Pbf}}J(\ol{\Pbf}^\star,\ul{\Pbf}^\star)+(\ul{\Pbf}^\star-\v{\ul{p}}_0)^\intercal\nabla_{\ul{\Pbf}}J(\ol{\Pbf}^\star,\ul{\Pbf}^\star)\nn\\
 %    &\geq \sum_{i=1}^N (\int_0^{\ol{P}^{i,\star}} \frac{\partial J(\cdot)}{\partial \ol{P}^i}d\ol{P}^i+\int_0^{\ul{P}^{i,\star}} \frac{\partial J(\cdot)}{\partial \ul{P}^i}d\ul{P}^i)\nn\\
     &\overset{(c)}{\geq} J(\ol{\Pbf}^\star,\ul{\Pbf}^\star)-J(\v{\ol{p}}_0,\v{\ul{p}}_0),
     \end{aligned}
\end{align}
rearranging which yields $\Pi^{\mbox{\tiny DSO}} \geq 0$, proving part (i).
The relation in (a) follows from Proposition \ref{Prop:LAP}. Inequality in (b) exploits the element-wise nonnegative nature of $\Abf_+, \Abf_-$, $\ol{\mubf}^\star, \ul{\mubf}^\star$,  $\ol{\omegabf}^\star, \ul{\omegabf}^\star$, $\ol{\Pbf}^\star-\v{\ol{p}}_0$, and $\ul{\Pbf}^\star-\v{\ul{p}}_0$. Convexity\footnote{A convex function $f$ has $f(\ybf)\geq f(\xbf)+\nabla f(\xbf)^\intercal (\ybf-\xbf)$.} of $J$ implies (c). 

$\bullet$ Part (ii) with the first condition: The KKT optimality conditions for \eqref{eq:auction.2} imply
\begin{gather}
\begin{gathered}
\nabla_{\v{\ol{C}}_k}{\cal L}^\star 
=-\nabla_{\v{\ol{C}}_k}\varphi_k(\v{\ul{C}}_k^\star,\v{\ol{C}}_k^\star)+\ol{\lambdabf}^\star - \ol{\etabf}^\star ={\bm 0},
\\
\nabla_{\v{\ul{C}}_k}{\cal L}^\star 
=-\nabla_{\v{\ul{C}}_k}\varphi_k(\v{\ul{C}}_k^\star,\v{\ol{C}}_k^\star)+\ul{\lambdabf}^\star- \ul{\etabf}^\star  ={\bm 0},
\\
\ol{\etabf}^{\star\intercal} \left( \v{\ol{C}}_k^{\mbox{\tiny min}} - \v{\ol{C}}_k^\star \right) = \ul{\etabf}^{\star\intercal} \left( \v{\ul{C}}_k^{\mbox{\tiny min}} - \v{\ul{C}}_k^\star \right) = 0
\end{gathered}
\label{eq:KKT.auction.2.2}
\end{gather}
with $\ul{\etabf}^\star \geq 0, \ol{\etabf}^\star \geq 0$. Using these relations, we infer
\begin{align}
\hspace{-0.1in}
\begin{aligned}
    \Pi_k^{\mbox{\tiny DERA}}
    &= \varphi_k(\v{\ul{C}}_k^\star,\v{\ol{C}}_k^\star) - \left(  \v{\ol{C}}_k^{\star\intercal} \ol{\lambdabf}^{\star} + \v{\ul{C}}_k^{\star\intercal} \ul{\lambdabf}^{\star} 
    \right)
    \\
    &=\varphi_k(\v{\ul{C}}_k^\star,\v{\ol{C}}_k^\star) 
    - \v{\ol{C}}_k^{\star\intercal} \nabla_{\v{\ol{C}}_k}\varphi_k(\v{\ul{C}}_k^\star,\v{\ol{C}}_k^\star)
    - \ol{\etabf}^{\star\intercal} \v{\ol{C}}_k^{\star}   
    \\
    & \quad 
    - \v{\ul{C}}_k^{\star\intercal} \nabla_{\v{\ul{C}}_k}\varphi_k(\v{\ul{C}}_k^\star,\v{\ol{C}}_k^\star)
    - \ul{\etabf}^{\star\intercal} \v{\ul{C}}_k^{\star}   
    \\
    &\overset{(a)}{\geq} \varphi_k(\v{0},\v{0}) 
    - \ol{\etabf}^{\star\intercal} \v{\ol{C}}_k^{\mbox{\tiny min}}   
    - \ul{\etabf}^{\star\intercal} \v{\ul{C}}_k^{\mbox{\tiny min}}
    \\
    &\overset{(b)}{\geq} 0,
\end{aligned}
\end{align}
where (a) utilizes the concavity of $\varphi_k$ and (b) utilizes $\varphi_k(\v{0},\v{0}) \geq 0$ and the fact that $\ol{\etabf}^{\star\intercal} \v{\ol{C}}_k^{\mbox{\tiny min}}   
    + \ul{\etabf}^{\star\intercal} \v{\ul{C}}_k^{\mbox{\tiny min}} = 0$ under either of the two hypotheses assumed for the result. Under the first hypothesis, $\v{\ol{C}}_k^{\mbox{\tiny min}} = \v{\ul{C}}_k^{\mbox{\tiny min}} = 0$ gives the result. Under the second, $ \ol{\etabf}^{\star\intercal} =  \ul{\etabf}^{\star\intercal} = 0$, owing to the last relation in \eqref{eq:KKT.auction.2.2}.

\vspace{-0.2in}
\subsection{Proof of Proposition~\ref{Prop:PM2}}\label{sec:ProfProp4}%Price monotonicity}
We only prove the result for $\ol{\lambdabf}^{\star}$; the proof for $\ul{\lambdabf}^{\star}$ is similar and omitted.
% The monotonic allocation price comes from the structure of  $\Abf$ for the LinDistFlow model. 
Recall the definition of $\v{A}$ from 
\eqref{eq:Abb.def}, for which all elements in $\v{A}$ are nonnegative. Note that all inequalities below represent element-wise relationships between two vectors. With bus $m$ as an ancestor of bus $n$,  we have $\tilde{\Sbf}(:,n) \geq  \tilde{\Sbf}(:,m)$, where the notation $\tilde{\Sbf}(:, j)$ identifies the column of $\tilde{\Sbf}$ corresponding to bus $j$. Recall that $\tilde{\Sbf}$ and $\v{D}$ are element-wise non-negative, implying that
\begin{align}
    \tilde{\Sbf}^\intercal \Dbf  \left[\tilde{\Sbf}(:,n) - \tilde{\Sbf}(:,m) \right] \geq \mathbf{0}.
\end{align}
In turn, this allows us to infer $\v{A}(:,n) \geq \v{A}(:,m) \geq \mathbf{0}$ and 
\begin{align}
    \v{A}(:,n)^{\intercal}\ol{\mubf}^\star \geq \v{A}(:,m)^{\intercal}\ol{\mubf}^\star,
    \label{eq:Amonoto}
\end{align}
because $\ol{\mubf}^\star \geq 0$. Furthermore, the KKT optimality conditions imply
$\ol{\omegabf}^{\star\intercal} \left( \ol{\Pbf}^{\star} - {\ol{\Pbf}}^{\mbox{\tiny max}} \right)=  0$, 
and
$\ol{\omegabf}^{\star} \geq 0$.
Under our hypothesis of the non-binding nature of \eqref{eq: ROPbd} at optimality, we infer $\ol{\omegabf}^{\star} = 0$.
Combining the observations in \eqref{eq:Amonoto} and $\ol{\omegabf}^{\star} = 0$ in Proposition \ref{Prop:LAP}, we then conclude
% \beq\label{eq:Amonoto}
% \abf^+_{n} \succeq \abf^+_{m} \Rightarrow  (\abf^+_{n})^{\T}\ol{\mubf}^\star \succeq (\abf^+_{m})^{\T}\ol{\mubf}^\star, 
% \eeq
% where $\abf_n$ is the $n$-th column of matrix $\Abf$, and $\abf_n=\abf^+_n-\abf^-_n$, representing the construction of $\abf_n$ out of its positive and negative entries. 
% 
% Therefore, we have
\begin{align}
% \hspace{-0.1in}
\begin{aligned}
\ol{\lambda}^{(m)\star} 
&= 
\ol{J} + \v{A}(:,m)^{\intercal}\ol{\mubf}^\star + \ol{\omega}^{(m)\star}
\\
&\leq
\ol{J} + \v{A}(:,n)^{\intercal}\ol{\mubf}^\star + \ol{\omega}^{(n)\star}
\\
&=
\ol{\lambda}^{(n)\star}
\end{aligned}
\end{align}
to complete the proof. We remark that the proof applies if $J(\ol{\Pbf},\ul{\Pbf}) = \sum_{i=1}^N  \ol{J}^{(i)} \ol{P}^{(i)} + \sum_{i=1}^N \ul{J}^{(i)} \ul{P}^{(i)}$ and they satisfy $\ol{J}^{(m)} \leq \ol{J}^{(n)}$, $\ul{J}^{(m)} \leq \ul{J}^{(n)}$.

% where (a) utilizes  $\Abf=\Abf^+$ and $\Abf_-={\bf 0}$ from \eqref{eq:Adef}, and KKT conditions (\ref{eq:KKTeq}). (b) utilizes the condition that the total injection access constraints (\ref{eq: ROPbd}) at each bus are nonbinding, which indicates  $\overline{\omegabf}^{\star}={\bf 0}$ by the complementary slackness condition. Additionally, (b) also uses the condition that DSO's cost is homogeneous over all buses, \ie $\frac{\partial J^\star(\cdot)}{\partial \ol{P}_{n}} =\frac{\partial J^\star(\cdot)}{\partial \ol{P}_{m}}$. Lastly, (c) utilizes the relation in \eqref{eq:Amonoto} and $\overline{\omegabf}^{\star}={\bf 0}$. 

% Similarly, we can prove $\ul{\lambda}^\star_m \le \ul{\lambda}^\star_{n}$ when $\frac{\partial J^\star(\cdot)}{\partial \ul{P}_{n}} =\frac{\partial J^\star(\cdot)}{\partial \ul{P}_{m}}$.

\vspace{-0.1in}
\subsection{Scenario-Approach for Stochastic Access Allocation}\label{sec:SO_Eq}

% Here we derive the CVaR-based security constraint and explain the reason why (\ref{eq:auction.CVaR}) can be approximated by  (\ref{eq:auction.3}).

The CVaR constraint (\ref{eq:CVaRPF}) and (\ref{eq:CVaRPF2}) in  (\ref{eq:auction.CVaR}) can be written equivalently as 
\begin{align}
& \forall \v{p}_k \in [-\v{\ul{C}}_k, \v{\ol{C}}_k],
\\
& \quad \begin{cases}
\mbox{CVaR}_\delta[\v{A}(\sum_{k=1}^K \v{p}_k + \v{p}_0) ]   \leq   \v{\ol{b}},
\\
\mbox{CVaR}_\delta[-\v{A}(\sum_{k=1}^K \v{p}_k + \v{p}_0) ] \leq -\v{\ul{b}},
\end{cases} 
\nn\\
& \overset{(a)}{\Leftrightarrow }
\forall \v{p}_k \in [-\v{\ul{C}}_k, \v{\ol{C}}_k],\nn\\
& ~~~~~~\begin{cases}
\v{A}\sum_{k=1}^K \v{p}_k + \mbox{CVaR}_\delta[\v{A} \v{p}_0]   \leq   \v{\ol{b}},
\\ 
-\v{A}\sum_{k=1}^K \v{p}_k +\mbox{CVaR}_\delta[-\v{A}\v{p}_0] \leq -\v{\ul{b}},
\end{cases} 
\nn\\
&   \overset{(b)}{ \Leftrightarrow}
    \begin{cases}
     \v{A}_+\sum_{k=1}^K \v{\ol{C}}_k+ \v{A}_-\sum_{k=1}^K \v{\ul{C}}_k
     \\
     \qquad \qquad - \v{\ol{b}}+ \mbox{CVaR}_\delta[\v{A}\v{p}_0 ]   \leq {\bf 0}, 
     \\
      \v{A}_-\sum_{k=1}^K \v{\ol{C}}_k+ \v{A}_+\sum_{k=1}^K \v{\ul{C}}_k
      \\
      \qquad \qquad +\v{\ul{b}}+ \mbox{CVaR}_\delta[-\v{A}\v{p}_0] \leq  {\bf 0}, \\
    \end{cases}\nn\\
&   \overset{(c)}{ \Leftrightarrow   }
    \begin{cases}
    \mbox{CVaR}_\delta[\v{A}_+\sum_{k=1}^K \v{\ol{C}}_k+ \v{A}_-\sum_{k=1}^K \v{\ul{C}}_k
    \\
     \qquad \qquad - \v{\ol{b}}+(\v{A}_+-\v{A}_-)\v{p}_0]   \leq {\bf 0}, 
     \\
     \mbox{CVaR}_\delta[\v{A}_-\sum_{k=1}^K \v{\ol{C}}_k+ \v{A}_+\sum_{k=1}^K \v{\ul{C}}_k
     \\
     \qquad\qquad +\v{\ul{b}}-(\v{A}_+-\v{A}_-)\v{p}_0] \leq  {\bf 0}, 
     \\
    \end{cases} \nn\\
    &    \overset{(d)}{\Leftrightarrow }  
    \begin{cases}
    \mbox{CVaR}_\delta[\v{A}_+ \v{\ol{P}}+ \v{A}_-\v{\ul{P}}- \v{\ol{b}}]   \leq {\bf 0}, \\
     \mbox{CVaR}_\delta[\v{A}_-\v{\ol{P}}+ \v{A}_+\v{\ul{P}}+\v{\ul{b}}] \leq  {\bf 0}, \\
    \end{cases} \label{eq:LV_CVaR1}
\end{align}
where (a) utilizes the property of CVaR that $\mbox{CVaR}_\delta[X+a]=\mbox{CVaR}_\delta[X]+a$ when $X$ is random and $a$ is a constant,  (b) is a consequence of the arguments in the proof of Lemma \ref{lemma:EqRO}, (c) uses $\Abf=\Abf_+-\Abf_-$, and (d) follows from the definition of $\ol{\Pbf}, \ul{\Pbf}$. 

Using \eqref{eq:CVarProbHold}, we rewrite \eqref{eq:LV_CVaR1} as
{\small
\begin{align*}
&\begin{cases}
\min_{\v{\ol{t}}} \big(\v{\ol{t}}+\frac{1}{S(1-\delta)}\sum_{s=1}^S[ \v{A}_+ \v{\ol{P}}[s]+ \v{A}_-\v{\ul{P}}[s]- \v{\ol{b}}-\v{\ol{t}}]_+\big)\le {\bf 0}, 
\\
\min_{\v{\ul{t}}} \big(\v{\ul{t}}+\frac{1}{S(1-\delta)}\sum_{s=1}^S[\v{A}_-\v{\ol{P}}[s]+ \v{A}_+\v{\ul{P}}[s]+\v{\ul{b}}-\v{\ul{t}}]_+\big)\le {\bf 0},\nn
\end{cases}
\\
&  \Leftrightarrow  \exists \v{\ol{t}},  \v{\ul{t}},\text{such that}
\\
&\begin{cases}
\v{\ol{t}}+\frac{1}{S(1-\delta)}\sum_{s=1}^S[ \v{A}_+ \v{\ol{P}}[s]+ \v{A}_-\v{\ul{P}}[s]- \v{\ol{b}}-\v{\ol{t}}]_+\le {\bf 0}, \\
\v{\ul{t}}+\frac{1}{S(1-\delta)}\sum_{s=1}^S[\v{A}_-\v{\ol{P}}[s]+ \v{A}_+\v{\ul{P}}[s]+\v{\ul{b}}-\v{\ul{t}}]_+\le {\bf 0},
\end{cases} 
\\
&  \Leftrightarrow  \exists \v{\ol{t}},  \v{\ul{t}}, \ol{\v{\gamma}}[s], \ul{\v{\gamma}}[s], s= 1,\ldots, S, \text{such that}
\\
&\begin{cases}
\v{\ol{t}}+\frac{1}{S(1-\delta)}\sum_{s=1}^S \ol{\v{\gamma}}[s] \leq {\bf 0},
\\
\v{A}_+ \v{\ol{P}}[s]+ \v{A}_-\v{\ul{P}}[s]- \v{\ol{b}}-\v{\ol{t}} \leq \ol{\v{\gamma}}[s],
\\
{\bf 0} \leq \ol{\v{\gamma}}[s],
\\
\v{\ul{t}}+\frac{1}{S(1-\delta)}\sum_{s=1}^S \ul{\v{\gamma}}[s] \leq {\bf 0},
\\
\v{A}_-\v{\ol{P}}[s]+ \v{A}_+\v{\ul{P}}[s]+\v{\ul{b}}-\v{\ul{t}} \leq \ul{\v{\gamma}}[s],
\\
{\bf 0} \leq \ul{\v{\gamma}}[s],
\end{cases}
\end{align*}
}
where
\begin{align}\label{eq:CVaReqSc}
 % \begin{cases} 
 \v{\ol{P}}[s]=\sum_{k=1}^K \v{\ol{C}}_k+\v{p}_0[s],
 % \\
 \;
\v{\ul{P}}[s]=\sum_{k=1}^K \v{\ul{C}}_k-\v{p}_0[s]
% \end{cases}
\end{align}
for $s = 1, \ldots, S$. This completes the derivation of \eqref{eq:auction.3}.
% , since (\ref{eq:SOepiol})-(\ref{eq:SOepiul}) expand  the “\_+” part into inequality constraints using the epigraph form.
% So we get the scenario-based constrains (\ref{eq:CVaR_RO}) and (\ref{eq:CVaReqSc}) that are used in  (\ref{eq:auction.3}).% Interestingly, if we begin with the robust formulation (\ref{eq:auction.2}) and find it's corresponding scenario-based stochastic program, we will also get constraints  (\ref{eq:CVaR_RO}) and (\ref{eq:CVaReqSc}).

\vspace{-0.2in}

\subsection{Proof of Proposition~\ref{Prop:LAP-SO} }\label{sec:ProfProp4}

% \tcb{
% \begin{proposition}\label{Prop:LAP-SO}
% LMAP-S satisfies
% \begin{align*}
% \ol{\Lambdabf}^\star
% &=\sum_{s=1}^S \frac{1}{S}{\nabla_{\ol{\Pbf}[s]}J(\ol{\Pbf}^{\star}[s],\ul{\Pbf}^{\star}[s])} 
% \\
% &\quad + \sum_{s=1}^S \left( \Abf_+^{\T} \ol{\betabf}^{\star}[s]+\Abf_-^{\T} \ul{\betabf}^{\star}[s] + \ol{\omegabf}^{\star}[s] \right),
% \\
% \ul{\Lambdabf}^\star
% &=\sum_{s=1}^S \frac{1}{S} {\nabla_{\ul{\Pbf}[s]}J(\ol{\Pbf}^{\star}[s]),\ul{\Pbf}^{\star}[s])}
% \\
% &\quad + \sum_{s=1}^S \left( \Abf_+^{\T} \ul{\betabf}^{\star}[s]+\Abf_-^{\T} \ol{\betabf}^{\star}[s] + \ul{\omegabf}^{\star}[s] \right).
% \end{align*}
% \end{proposition}
% }
Denote $J^{\star}[s]:=J(\ol{\Pbf}^{\star}[s],\ul{\Pbf}^{\star}[s])$. The KKT optimality conditions for \eqref{eq:auction.3} yield
\begin{align}\label{eq:KKTCVaR}
     \ol{\lambdabf}^{\star}[s]= \frac{1}{S}\nabla_{\ol{\Pbf}[s]}J^{\star}[s]+\Abf_+^{\T}\ol{\betabf}[s]+\Abf_-^{\T}\ul{\betabf}^{\star}[s]+\ol{\omegabf}^{\star}[s],
    \nn\\
     \ul{\lambdabf}^{\star}[s] =\frac{1}{S}\nabla_{\ul{\Pbf}[s]}J^{\star}[s]+\Abf_+^{\T}\ul{\betabf}[s]+\Abf_-^{\T}\ol{\betabf}^{\star}[s]+\ul{\omegabf}^{\star}[s].
\end{align}
The rest of part (a) follows from summing the above across scenarios with $\ol{\Lambdabf}^\star = \sum_{s=1}^S \ol{\lambdabf}^{\star}_s$ and $\ul{\Lambdabf}^\star = \sum_{s=1}^S \ul{\lambdabf}^{\star}_s$.
Proofs of parts (b) and (c) follow upon replacing $(\ol{\lambdabf}, \ul{\lambdabf})$ with $(\ol{\Lambdabf}, \ul{\Lambdabf})$, and $J(\ol{\Pbf}^\star,\ul{\Pbf}^\star)-J(\v{\ol{p}}_0,\v{\ul{p}}_0)$ with $\frac{1}{S}\sum_{s=1}^SJ(\ol{\Pbf}^{\star}[s],\ul{\Pbf}^{\star}[s])-\frac{1}{S}\sum_{s=1}^SJ(\v{\ol{p}}_0[s],\v{\ul{p}}_0[s])$ in the proof of Proposition \ref{Prop:RA}.
Part (d) follows from the proof of Proposition~\ref{Prop:PM2}, upon replacing $(\ol{\lambdabf}, \ul{\lambdabf})$ with $(\ol{\Lambdabf}, \ul{\Lambdabf})$.

\subsection{DERAs' Bid-In Utility Functions for the 141-Bus System}
\label{sec:paper2.141}
To participate in the network access allocation, we assumed that all four DERAs adopted the aggregation method in \cite{ChenAlahmedMountTong23DERA} of this paper series to submit the bid-in utility functions $\varphi_k$ and the  minimum network access limits. The bid-in parameters for the four DERAs are shown in Table \ref{table:benefit141}. They were computed from equation (9) of  \cite{ChenAlahmedMountTong23DERA}, which in turn is derived from the DERA's profit maximization problem. We consider competitive DER aggregation that maximizes the DERA's profit, subject to its customers gaining higher surpluses than what a regulated utility company can offer.

As for the parameter settings for equation (9) in \cite{ChenAlahmedMountTong23DERA}, we used $\zeta=1.01$, $\pi^0=\$0$, $\pi^+=\$0.3$/kWh, and $\pi^{-}=\$0.12$/kWh  \cite{PGE22NEM} for the NEM X tariff, and $\pi_{\mbox{\tiny  LMP}}=\$0.1$/kWh \cite{CAISO22price} for the wholesale market LMP. Additionally, each DERA aggregated 50 households on the buses they aggregated. Homogeneous quadratic utility functions  were used for the households with $\hat{a}=\$0.4/\mbox{kWh}, \hat{b}=\$0.1/(\mbox{kWh})^2$ \cite{SamadiSchoberWong12TSG} in $$U(x)=\begin{cases}
\hat{a} x-\frac{\hat{b}}{2}x^2,&0\le x\le \frac{\hat{a}}{\hat{b}},\\ 
\frac{\hat{a}^2}{2\hat{b}},&x> \frac{\hat{a}}{\hat{b}}.
\end{cases}$$ 
As shown in Fig. 6 of  \cite{ChenAlahmedMountTong23DERA}, a DERA with a higher DG level had a higher marginal surplus for injection access. Correspondingly, in our result (Fig.~\ref{fig:MC}), we observe that DERA 3 and DERA 4 with higher BTM DG generation acquired more injection access. Buses  118-134 with DERA 4 exhibited the highest $\ol{\lambda}_i^\star$ because DERA 4 had the higher BTM DG generation and thus, the higher incentive to purchase injection access, compared to DERA 3.
\end{document}